\documentclass[twoside,twocolumn,9pt]{article}
\usepackage{ulem}
\usepackage{extsizes}
\usepackage[super,sort&compress,comma]{natbib} 
\usepackage[version=3]{mhchem}
\usepackage[left=1.5cm, right=1.5cm, top=1.785cm, bottom=2.0cm]{geometry}
\usepackage{balance}
\usepackage{widetext}
\usepackage{times,mathptmx}
\usepackage{sectsty}
\usepackage{graphicx} 
\usepackage{lastpage}
\usepackage[format=plain,justification=raggedright,singlelinecheck=false,font={stretch=1.125,small,sf},labelfont=bf,labelsep=space]{caption}
\usepackage{float}
\usepackage{fancyhdr}
\usepackage{fnpos}
\usepackage[english]{babel}
\usepackage{array}
\usepackage{droidsans}
\usepackage{charter}
\usepackage[T1]{fontenc}
\usepackage[usenames,dvipsnames]{xcolor}
\usepackage{setspace}
\usepackage[compact]{titlesec}
\usepackage{latexsym}
\usepackage{subcaption}

\definecolor{cream}{RGB}{222,217,201}

\makeFNbottom
\makeatletter
\renewcommand\LARGE{\@setfontsize\LARGE{15pt}{17}}
\renewcommand\Large{\@setfontsize\Large{12pt}{14}}
\renewcommand\large{\@setfontsize\large{10pt}{12}}
\renewcommand\footnotesize{\@setfontsize\footnotesize{7pt}{10}}
\makeatother

\renewcommand\footnoterule{\vspace*{1pt}%
\color{cream}\hrule width 3.5in height 0.4pt \color{black}\vspace*{5pt}} 
\makeatletter 
\renewcommand\@biblabel[1]{#1}            
\renewcommand\@makefntext[1]%
{\noindent\makebox[0pt][r]{\@thefnmark\,}#1}\makeatother 
\renewcommand{\figurename}{\small{Fig.}~}
\sectionfont{\sffamily\Large}
\subsectionfont{\normalsize}
\subsubsectionfont{\bf}
\titlespacing*{\section}{0pt}{4pt}{4pt}
\titlespacing*{\subsection}{0pt}{15pt}{1pt}

\renewcommand{\headrulewidth}{0pt} 
\renewcommand{\footrulewidth}{0pt}
\makeatletter 
\newlength{\figrulesep} 
\makeatother

\newcommand{\trialfrequency}{$1.0\times10^{12}$ s$^{-1}$} 
\newcommand{\protonenergy}{100 } 
\newcommand{\protonflux}{$1.0\times10^{11}\;\mathrm{cm}^{-2}\;\mathrm{s}^{-1}$}
\newcommand{\rhoIce}{$1.313\times10^{22}\;\mathrm{cm}^{-3}$}

\newcommand{\flux}{$10^{11}\;\mathrm{cm}^{-2}\;\mathrm{s}^{-1}$ }

\title{\sffamily \textbf{A New Model of the Chemistry of Ionizing Radiation In Solids: CIRIS}}

\author{\sffamily \textbf{Christopher N. Shingledecker$^{\ast}$\textit{$^{a}$}, Romane Le Gal$^\textit{a,b}$, and Eric Herbst\textit{$^{a,b}$}}}

\begin{document}
\maketitle

\begin{abstract}
        The collisions between high-energy ions and solids can result in
        significant physical and chemical changes to the material. These
        effects are potentially important for better understanding the
        chemistry of interstellar and planetary bodies, which are exposed to
        cosmic radiation and the solar wind, respectively; however, modeling
        such collisions on a detailed microscopic basis has thus far been
        largely unsuccessful. To that end, a new model, entitled
        \texttt{CIRIS}: the Chemistry of Ionizing Radiation in Solids, was
        created to calculate the physical and chemical effects of the
        irradiation of solid materials. With the new code, we simulate
        O$_2$ ice irradiated with 100 keV protons. Our models are able to
        reproduce the measured ozone abundances of a previous experimental
        study, as well as independently predict the approximate
        thickness of the ice used in that work.
\end{abstract}


\footnotetext{\textit{$^{\ast}$~Corresponding author: P.O. Box 400319, Charlottesville, VA 22904, USA. Tel: 434 831 6240; E-mail: shingledecker@virginia.edu}}
\footnotetext{\textit{$^{a}$~Department of Chemistry, University of Virginia, Charlottesville, VA 22904, USA}}
\footnotetext{\textit{$^{b}$~Department of Astronomy, University of Virginia, Charlottesville, VA 22904, USA}}

\section{Introduction}

Irradiation by charged particles is well-known to cause substantial
physiochemical changes in condensed matter.  Beyond its more obvious
connections to medical and material science, a detailed understanding of the
effects of irradiation induced processes is of great astrochemical interest,
since the galaxy is bathed in cosmic rays. Cosmic rays are a particularly
high-energy form of ionizing radiation (MeV - TeV) \citep{ikha1991} comprised
mostly of protons, which are thought to be created in supernovae
\citep{baade1934} or by the super-massive black holes at the centers of
galaxies \citep{abraham2007}. In the interstellar medium, cosmic rays are known
to have a strong influence on the chemistry of molecular clouds
\citep{grenier2015}. One component of such environments is dust, which, in cold
dense regions, is covered by an ice mantle mainly composed of water
\citep{hasegawa1992,herbst2009}. Based on the substantial body of experimental
studies showing the complexity of irradiation chemistry, such processes
represent promising means by which large interstellar molecules could be formed
\citep{abplanalp2016}, particularly in cases where observational results are
difficult to reproduce with current astronomical models \citep{corby2015} and
possible drivers of complex molecule synthesis are not obvious. In the case of
interstellar clouds such as TMC-1, which have temperatures of around 10 K and
densities of $\sim 10^4$ cm$^{-3}$, a better understanding of cosmic ray
induced irradiation chemistry is particularly required since the interiors of
these regions are shielded from most of the interstellar UV radiation field.
Though cosmic ray ionization rates are reduced by roughly two orders of
magnitudes in these regions \citep{rimmer2012},  a steady-state is reached at
which ionization rates stabilize. Thus, particularly in dense cold interstellar
environments, it is possible that cosmic ray induced chemical processes
represent a potentially very efficient pathway to produce complex and even
prebiotic molecules.

Although the interstellar chemistry following cosmic ray bombardment of grains
has rarely been studied theoretically, Monte Carlo techniques have been used to
give a detailed, microscopic view of the chemistry of solids with, however,
only a cursory treatment of the role of ion bombardment. The first such
stochastic grain chemistry model utilizing a Monte Carlo approach was reported
by \citet{chang2005}. This and later  models have utilized in particular the
continuous-time random-walk Monte Carlo approach of \citet{montroll1965}, such
as the simulation of \citet{chang2014}.

Due to its many practical applications, prior interest in modeling irrradiated
matter is well known. Most of this previous computational research has focused
on simulating particle tracks in the material, which result from the collisions
between species in the solid and the irradiating particles.  Due to the
stochastic nature of these collisions, Monte Carlo methods are a natural choice
for such simulations.  Well-known models of this type include \texttt{MOCA}
\citep{paretzke1974}, \texttt{MARLOWE} \citep{robinson1974}, \texttt{TRIM}
\citep{ziegler1988}, and that of \citet{pimblottlavernemozumder}.  The
subsequent irradiation-induced chemistry that occurs has been followed in a
large number of laboratory experiments on ices and bare solids using both
high-energy protons and electrons. In spite of this interest, there has been
only limited success in combining track calculations with the subsequent
chemistry.  One of the most detailed of such attempts was the model of
\citet{pimblottlaverne2002}. In that work, the authors were able to combine
realistic track calculations with a simplified chemical network representing
the aqueous solution of a Fricke dosimeter; however, in spite of approximations
to the chemistry, such as treating the solute as an infinite continuum, they
were only able to simulate the radiation induced chemistry for $\sim 1 \mu$s,
due to the computational expense.

The computational expense of these models reflects the many complex physical
and chemical processes associated with the bombardment of solids, which must be
considered when using a detailed microscopic approach like molecular dynamics
or one of the Monte Carlo methods. The physical processes are initiated when a
moving energetic particle collides with some material, which we call the
target. In this work, the particles we consider are protons and we refer to
these as primary ions; however, if the incoming particle is an electron it is
sometimes called the primary electron.  These primaries transfer energy to
atomic or molecular species in the target through collisions. Some of these
collisions ionize species in the material, resulting in the formation of
``secondary'' electrons, which, in turn, transfer energy collisionally to
target species and compound the effects of the primary particle, in large part
by causing the formation of additional charged species and secondary electrons
\citep{johnson1990}.  Subsequent charge recombination reactions help drive the
formation of radicals and other highly reactive species \citep{mason2014},
which can react via thermal diffusive mechanisms in the solid, albeit much more
slowly than the non-thermal irradiation induced processes \citep{johnson1990}.

In this paper, we present a code that is designed to simulate these processes
over simulated irradiation exposures relevant to experiments and other
real-world applications where a better understanding of the resulting effects
on the material is desired. This program, with the acronym \texttt{CIRIS},
which stands for the Chemistry of Ionizing Radiation in Solids, represents an
initial attempt at the development of a simulation including a unified model of
atomic physics and chemistry. Here, we report the use of our code to simulate
the chemistry of the experimental system studied by \citet{baragiola1999},
hereafter referred to as B99. In that work, solid O$_2$ ice cooled to 5 K under
ultra-high vacuum (UHV) conditions was irradiated with 100 keV protons and
ozone was synthesized via proton-induced chemical processes. 

This system is particularly well-suited as an initial test of our new code. Due
to the novelty of this kind of model, well-constrained experiments such as the
one reported by B99 allow for a reasonable comparison with our theoretical
data.  This comparison is aided by the relative simplicity of the irradiated
oxygen ice system. One example of this relative simplicity is the upper limit
to the size of observable molecules produced via irradiation; namely, ozone, as
found by experimental studies, not only of B99, but also those of
\citet{ennis2012} and \citet{lacombe1997}. In other systems, such as those
containing carbon, it may not be obvious to determine the limit of chemical
complexity obtainable through these processes and some arbitrary upper limit in
molecular size may have to be set in the chemical network.

Molecular oxygen ice is present, not only in the Solar System on icy Jovian
moons such like Ganymede \citep{spencer1995,calvin1997} but also in comets.
Recently, interest in cometary O$_2$ ice has increased following the detection
of gas-phase O$_2$ around comet 67P/Churyumov-Gerasimenko \citep{bieler2015}.
It has been speculated by \citet{taquet2016} and \citet{mousis2016} that this
molecular oxygen was liberated from the icy body of the comet. If that is
indeed true, then this represents a possible immediate application of the
current results, since comets experience irradiation from the Solar wind, made
up mostly of protons with energies mostly between 1.5 and 10 keV
\citep{schwenn2001}, and cosmic rays with energies above $\sim$1 MeV
\citep{spitzer1968} that are not stopped by the Solar wind. Moreover, it is
possible that cometary ices are relatively pristine remnants of the parent
pre-solar nebula \citep{taquet2016}.  Since observations of interstellar O$_2$,
either in the gas or frozen in ice, have thus far proved mainly unsuccessful
\citep{goldsmith2000,pagani2003,yildiz2013}, cometary ice chemistry may provide
a crucial window into an as yet poorly understood aspect of interstellar
chemistry. Therefore, it is possible that a better understanding of the
irradiation chemistry of cometary ices in our Solar system can provide clues as
to its possible importance in and contribution to more remote interstellar
environments.

The format of the following sections of the paper is as follows. In Section 2
we discuss our model in more detail and give the theory behind the atomic
physics  and chemistry calculations. In Section 3, we give the results and
discuss how these compare with prior experimental work. Finally, in Section 4,
we present our summary and conclusions. 

\section{Model and Theory}

In our approach, solids are approximated as a lattice, represented in the
program by a three-dimensional array. This structure is comprising two types of
sites: strongly bound regular lattice sites on the surface and in the bulk of
the solid, and more weakly bound internal interstitial sites
\citep{akiyama1987}.  At the start of a simulation, we assume a regular lattice
comprised of some material, which in this work is solid O$_2$. The simulation
begins when the first irradiating proton collides with the pristine ice.  The
primary ion strikes a random surface site at a 90$^\circ$ angle and, as it
travels through the solid, the code calculates the relevant physical changes to
the material, as described below in Section \ref{rad-phase}.  The calculations
described there are repeated for every subsequent particle arrival. As given in
Section \ref{chem-phase}, in the model times between particle arrivals, neutral
species can thermally diffuse through the solid and chemical reactions in
Section \ref{network} can occur.  The model ends when the target material has
been exposed to some total amount of irradiation, known as the fluence.

\subsection{Monte Carlo modeling of irradiation effects} \label{rad-phase}

Collisions between energetic particles, such as protons and electrons, with
target species occur randomly along the track of the particle through the solid
and are thus well-suited to modeling using Monte Carlo techniques. For
so-called ``fast'' incident ions with energies greater than $\sim$ 1 keV/amu
\citep{johnson1990}, the timescale for these collisions is very short, relative
to the chemical timescale \citep{johnson1990}. Because the physical irradiation
processes occur much faster than the subsequent chemistry, the model decouples
the track calculations from the chemistry while the ion is travelling through
the target.

In our code, the arrival rate, in $s^{-1}$, for incoming primary ions is given
by

\begin{equation}
  k_\mathrm{ion} = \phi_\mathrm{ion}A_\mathrm{solid}
\end{equation}

\noindent 
where $\phi_\mathrm{ion}$ is the radiation flux in cm$^{-2}$ s$^{-1}$ and
$A_\mathrm{solid}$ is the surface area of the solid being irradiated.  We
assume a Poisson distrubution of waiting times and calculate the next ion
arrival using the stochastic relation

\begin{equation}
  \tau_\mathrm{ion} = -\frac{\mathrm{ln}(R_\mathrm{n})}{k_\mathrm{ion}}
\end{equation}

\noindent
where $R_\mathrm{n}$ is a pseudorandom number between 0 and 1.  When an ion
hits the ice, thermal diffusion is halted until a special set of calculations,
described in section \ref{track_calculations}, is completed to determine the
changes to the solid caused by the incoming ion.

\subsubsection{Track Calculations} \label{track_calculations}

Following \citet{bohr1913}, we divide collisions between a moving energetic ion
and stationary target into two broad categories: those in which energy is
transferred to the target nuclei, which are customarily called nuclear, or
elastic, collisions, and those in which energy is transferred to the target
electrons, which are called electronic, or inelastic, collisions. The inelastic
collisions can be further subdivided into ionizations, in which enough energy
is imparted to liberate a secondary electron, and electronic excitations of the
target atom or molecule. Here, rotational and vibrational excitations are not
considered.

Our code calculates the distance to the next collision, $\Delta x$, based on
the mean-free-path, $\Lambda_\mathrm{tot}$ for an energetic ion with some
energy, $E$, given by

\begin{equation}
  \Lambda_{\mathrm{tot}} = \frac{1}{n_{\mathrm{target}}\,\sigma_{\mathrm{tot}}}
    \label{track-mfp}
\end{equation}

\noindent
where $n_{\mathrm{target}}$ is the number density of the target and
$\sigma_{\mathrm{tot}}$ is the total cross-section
\citep{pimblottlavernemozumder}, which is given by

\begin{equation}
  \sigma_{\mathrm{tot}} = \sigma_{\mathrm{elastic}} + \sigma_{\mathrm{inelastic}}  = \sigma_{\mathrm{elastic}} + \sigma_{\mathrm{ion}} + \sigma_{\mathrm{ex}}
    \label{total-sigma}
\end{equation}

\noindent
where $\sigma_{\mathrm{ion}}$ and $\sigma_{\mathrm{ex}}$ are the cross-sections
for ionization and excitation of the target, respectively
\citep{pimblottlavernemozumder}. We stochastically determine which of these
three types of collisions occurs next based on the relative sizes of the
cross-sections using:

  \begin{eqnarray} \label{collision-scheme}
    \text{Collision}
    \begin{cases}
      & \text{Ionization for } 0 < R_\mathrm{n} \leq \frac{\sigma_\mathrm{ion}}{\sigma_\mathrm{tot}}\\
      & \text{Excitation for } \frac{\sigma_\mathrm{ion}}{\sigma_\mathrm{tot}} < R_\mathrm{n} \leq \frac{\sigma_\mathrm{inelastic}}{\sigma_\mathrm{tot}}\\
      & \text{Elastic for } \frac{\sigma_\mathrm{inelastic}}{\sigma_\mathrm{tot}} < R_\mathrm{n} \leq 1.
    \end{cases}
  \end{eqnarray}

\noindent
To obtain the distance to the next collision event, we sample from another
Poisson distribution using

\begin{equation}
  R_\mathrm{n} = 1 - e^{-\Delta x / \Lambda_{\mathrm{tot}}}
  \label{track.eps}
\end{equation}

\noindent
where here, a different random number, $R_\mathrm{n}$, is used and is
equivalent to the probability of the particle travelling a distance of $\Delta
x$ to the next collision, given a mean-free-path of $\Lambda_\mathrm{tot}$
\citep{pimblottlavernemozumder}.

\subsubsection{Protons}

For protons, we calculate the three key cross-sections for elastic collisions,
ionizations and excitations, given in eq. \eqref{total-sigma} before the first
collision, and again after every subsequent one, until the particle leaves the
system or its energy falls below an arbitrary cutoff of 100 eV, which is the
lower limit to the applicability of our method \citep{biersack1980}. The
average energy loss per unit path length of an ion, labelled A, in a material
made of either atoms or molecules, labelled B, is called the stopping, or
stopping power \citep{johnson1990}, and is given, in the laboratory coordinate
system, by

\begin{equation}
  \frac{dE}{dx} = n_\mathrm{B}(S_\mathrm{n}(E) + S_\mathrm{e}(E))
  \label{stopping}
\end{equation}

\noindent
where  $S_\mathrm{n}$ and $S_\mathrm{e}$ are the so-called stopping
cross-sections for nuclear (elastic) and electronic (inelastic) collisions,
respectively, in units of $\mathrm{eV}\;\mathrm{cm}^2$, and $n_\mathrm{B}$ is
the number density of the material. Here, A refers to protons only, and where B
is a molecule, we follow Bragg's rule in approximating the stopping powers as
linear combinations of the stopping powers of its constituent atoms
\citep{ziegler1985, ziegler1988}. 

\begin{figure}[htb!]
  \centering
  \begin{subfigure}[t]{0.48\textwidth}
    \includegraphics[height=5.5cm]{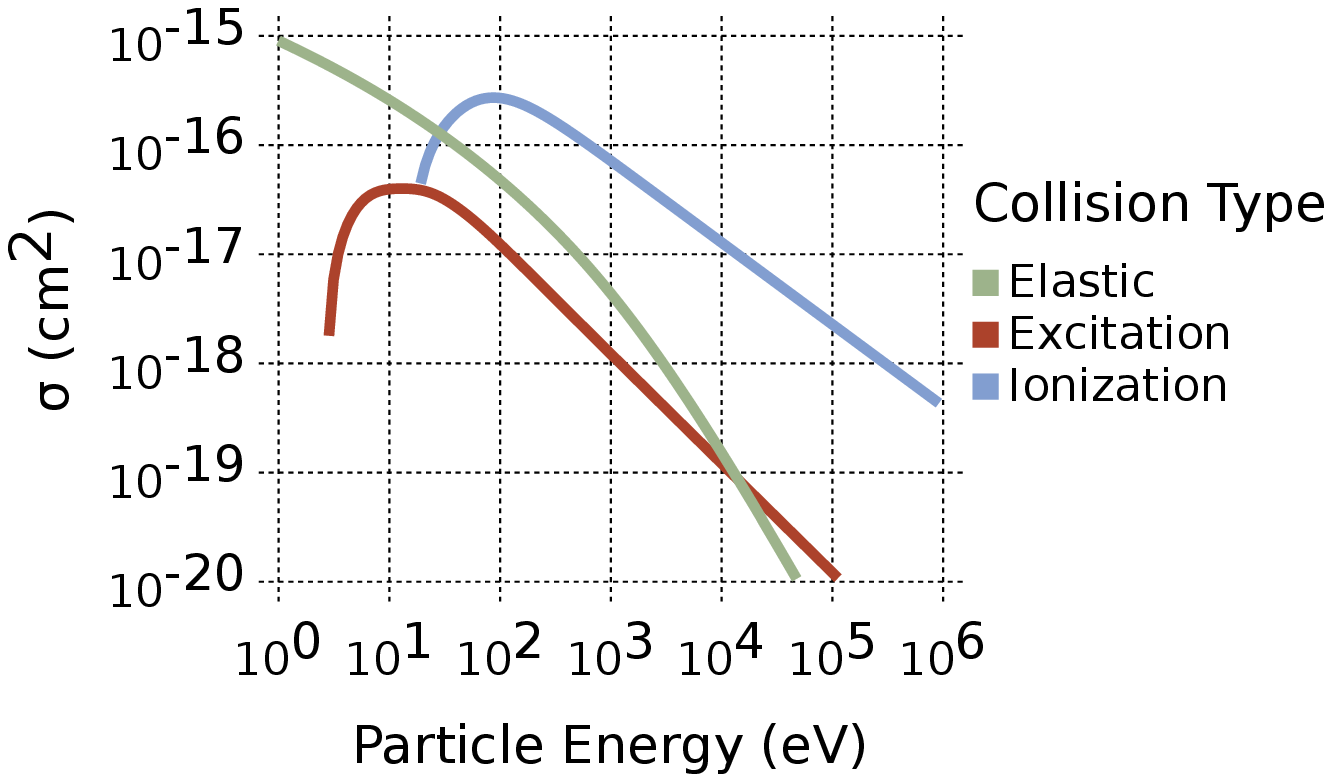}
    \caption{Atomic oxygen}
    \label{f1a}
  \end{subfigure}
  \begin{subfigure}[t]{0.48\textwidth}
    \includegraphics[height=5.5cm]{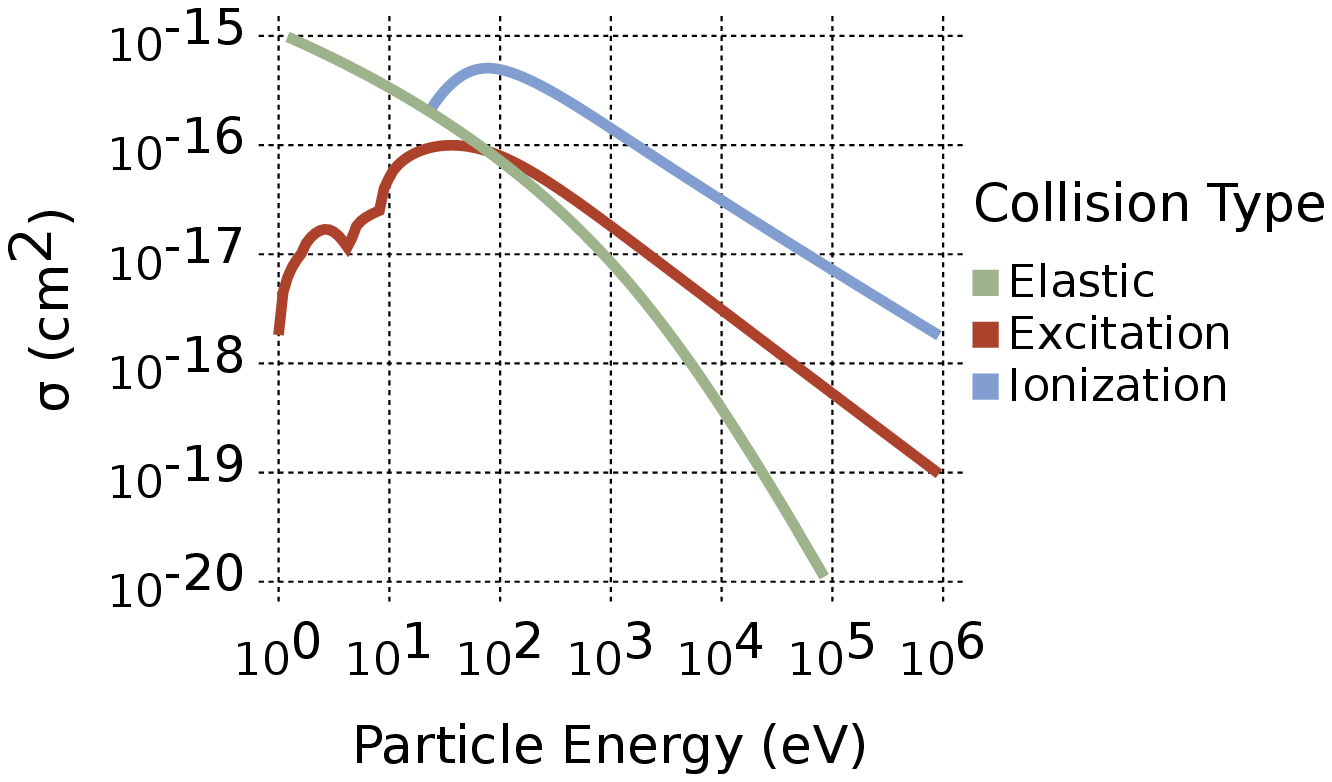}
    \caption{Molecular oxygen}
    \label{f1b}
  \end{subfigure}
  \begin{subfigure}[t]{0.48\textwidth}
    \includegraphics[height=5.5cm]{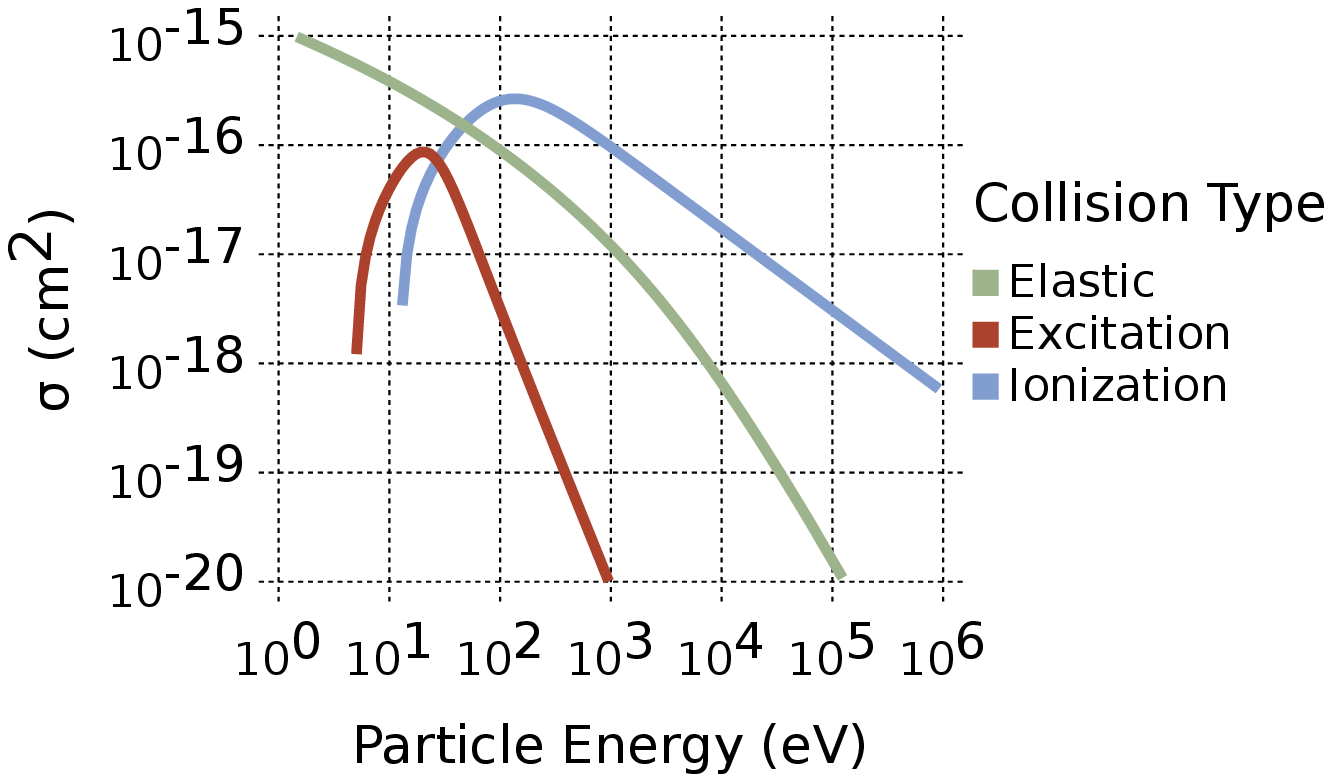}
    \caption{Ozone}
    \label{f1c}
  \end{subfigure}
  \caption{Proton-collision cross-sections using parameters listed in Tables \ref{proton_ionization_parameters} and \ref{proton_excitation_parameters}.}
  \label{f1}
\end{figure}

Our code uses the stopping cross-section, $S_\mathrm{n}$, to calculate the
collisional cross-section, $\sigma_\mathrm{elastic}$. There are many
semi-empirical expressions for calculating $S_\mathrm{n}$ \citep{johnson1990};
we utilize the formalism developed by \citet{ziegler1985} and used in the
\texttt{TRIM} program \citep{ziegler1985,ziegler1988} which is valid for ions
with energies between $\sim$0.1 keV and several MeV. For an elastic collision
between ion A, with energy, $E_\mathrm{A}$, and atom B, this is given by

\begin{equation}
  S_\mathrm{n}(E) = 2\pi \frac{\mathcal{A}^2}{\gamma E_{\mathrm{A}}}[2\epsilon\,s_{\mathrm{n}}(\epsilon)].
  \label{proton-nuclear-stopping-cross-section}
\end{equation}

\noindent
Here, $\gamma$ is the mass fraction, defined as

\begin{equation}
  \gamma = \frac{4m_{\mathrm{A}}m_{\mathrm{B}}}{(m_{\mathrm{A}}+m_{\mathrm{B}})^2},
  \label{mass-fraction}
\end{equation}

\noindent
and $\mathcal{A}$ is a quantity given by

\begin{equation}
  \mathcal{A} = \left( \frac{2m_{\mathrm{A}}}{m_{\mathrm{A}}+m_{\mathrm{B}}} Z_{\mathrm{A}}Z_{\mathrm{B}}e^2 \right)
\end{equation}

\noindent
where $Z_{\mathrm{A}}$ and $Z_{\mathrm{B}}$ are the nuclear charges of A and B.
In equation \eqref{proton-nuclear-stopping-cross-section}, $\epsilon$ is the
Lindhard-Scharff-Schiott (LSS) reduced energy \citep{lindhard1963}, calculated
as

\begin{equation}
  \epsilon = \frac{32.53\,m_\mathrm{B}\,E_\mathrm{A}}{Z_\mathrm{A}\,Z_\mathrm{B}\,(m_\mathrm{A}+m_\mathrm{B})(Z_\mathrm{A}^{0.23}+Z_\mathrm{B}^{0.23})}
\end{equation}

\noindent
and $s_n(\epsilon)$ is the reduced stopping, which has a value of

  \begin{eqnarray} \label{LSS-stopping}
    s_n(\epsilon)=
    \begin{cases}
      & \frac{\mathrm{ln}(\epsilon)}{2\,\epsilon} \;\; \text{when} \;\; \epsilon > 30,\\
      \\
      & \frac{\mathrm{ln}(1 + 1.1383\epsilon)}{(\epsilon \; + \; 0.01321\epsilon^{0.21226} \;+\; 0.19593\epsilon^{0.5})} \;\; \text{otherwise}.
    \end{cases}
  \end{eqnarray}

\noindent
We thus calculate the elastic collisional cross-section from the stopping
cross-section using

\begin{equation}
  \sigma_\mathrm{elastic} = \frac{2S_\mathrm{n}(E)}{\gamma E_\mathrm{A}}.
\end{equation}

\noindent
These elastic cross-sections typically have values of less than $\sim10^{-20}$
cm$^{-2}$ at energies above approximately 100 keV and approach values of
$10^{-16}$ cm$^{-2}$ as the incident ion energy decreases.

The kinetic energy transferred from the ion to the target in the elastic
collision is found from elementary classical scattering theory
\citep{ziegler1988} to be

\begin{equation}
  E_\mathrm{kin} = \frac{4E_\mathrm{A}m_\mathrm{A}m_\mathrm{B}}{m_\mathrm{A} + m_\mathrm{B}}\mathrm{sin^2}\frac{\Theta}{2}
\end{equation}

\noindent
where $\Theta$, the center-of-mass (CM) scattering angle, is obtained using the
``Magic Formula'' of \citet{biersack1980} and the ``universal'' potential for
ion-target scattering of \citet{ziegler1985}. It should be noted that the
scattering angles are all assumed to be small and thus tracks of the primary
ions are approximated with straight line trajectories; however, we calculate
these angles explicitly in the code to allow for more accurate modeling of the
energy lost in such nuclear-elastic collisions.

  \begin{table}[htb]
    \small
    \caption{Proton ionization cross-section parameters}
    \label{proton_ionization_parameters}
    \begin{tabular*}{0.5\textwidth}{@{\extracolsep{\fill}}cccccc}
      \hline
      State & $a$ & $J$(eV) & $\nu$ & $\Omega$ & $I$(eV) \\
      \hline
      \multicolumn{6}{c}{Atomic Oxygen\footnotemark[1]} \\
      $^3\mathrm{P}_2$ & 20.40 & $6.15\times10^4$ & 0.82 & 0.75 & 13.6 \\
      \hline
      \multicolumn{6}{c}{Molecular Oxygen\footnotemark[1]} \\
      $X\;^2\Pi_g$ & 9.56 & $4.58\times10^4$ & 0.61 & 1.26 & 12.1 \\
      \hline
      \multicolumn{6}{c}{Ozone\footnotemark[2]} \\
      $X\;^1\mathrm{A}_1$ & 40.00 & $1.05\times10^5$ & 1.00 & 0.75 & 12.43 \\
      \hline 
    \end{tabular*}
  \end{table}
  
\footnotetext[1]{Data taken from \citet{edgar1975}}
\footnotetext[2]{Data extracted from \citet{newson1995}}

  \begin{table}[htb]
    \small
    \caption{Proton excitation cross-section parameters}
    \label{proton_excitation_parameters}
    \begin{tabular*}{0.5\textwidth}{@{\extracolsep{\fill}}cccccc}
      \hline
      Excited State & $a$ & $J$(eV) & $\nu$ & $\Omega$ & $W$(eV) \\
      \hline
      \multicolumn{6}{c}{Atomic Oxygen\footnotemark[1]} \\
      $^1\mathrm{D}$ & 0.51 & $5.40\times10^3$ & 1.0 & 1.0 & 1.85 \\
      $^1\mathrm{S}$ & 0.075 & $8.70\times10^3$ & 1.0 & 1.0 & 4.18 \\
      $^3\mathrm{S}$ & 0.38 & $3.72\times10^4$ & 1.0 & 1.0 & 9.53 \\
      $^5\mathrm{S}$ & 0.55 & $1.61\times10^6$ & 1.0 & 1.0 & 9.20 \\
      \hline
      \multicolumn{6}{c}{Molecular Oxygen\footnotemark[1]} \\
      $a^1\Delta_\mathrm{g}$ & 0.092 & $2.50\times10^3$ & 0.5 & 3.0 & 0.98 \\
      $b^1\Sigma_\mathrm{g}^+$ & 0.11 & $4.19\times10^3$ & 0.5 & 3.0 & 1.64 \\
      $A^3\Sigma_\mathrm{u}^+$ & 0.57 & $1.76\times10^4$ & 0.5 & 0.9 & 4.5 \\
      $B^3\Sigma_\mathrm{u}^-$ & 4.73 & $5.17\times10^4$ & 0.5 & 0.75 & 8.4 \\
      9.9 eV peak & 0.83 & $8.07\times10^4$ & 0.5 & 0.85 & 9.9 \\
      \hline
      \multicolumn{6}{c}{Ozone\footnotemark[3]} \\
      $^1\mathrm{B}_2$ & 0.84 & 20.00 & 0.30 & 35.00 & 4.9 \\
      \hline
    \end{tabular*}
  \end{table}

\footnotetext[3]{Data extracted from \citet{sweeney1996}}

For protons, ionization and excitation cross-sections are calculated using the
semi-empirical Green-McNeal formula \citep{green1971,miller1973}

\begin{equation}
  \sigma_\mathrm{inelastic} = \sigma_0 \frac{(Z_\mathrm{B}\,a)^\Omega(E_\textrm{A}-I)^\nu}{J^{\Omega + \nu}\;+\;(E_\textrm{A}-I)^{\Omega + \nu}}
  \label{green-mcneal-ion}
\end{equation}

\noindent
where $\sigma_0 = 10^{-16}\approx \pi\,r_0^2$ cm$^2$, $Z_\mathrm{B}$ is the
number of electrons in the target atom or molecule, and $I$ is the ionization
threshold. For excitations between bound states, we replace $I$ with $W$, the
energy threshold for the particular transition.  Listed in Tables
\ref{proton_ionization_parameters} and \ref{proton_excitation_parameters} are
the species-dependent values of the unitless parameters $\Omega$, $\nu$, and
$a$, as well as those of the parameter $J$, which is given in units of energy.  

In the O$_2$ ice system considered here, ionization of neutrals by protons, as
well as secondary electrons, leads to the formation of the following
cation-secondary electron product pairs:

\begin{equation}
  \mathrm{O} \rightarrow \mathrm{O}^+ + e^- 
\end{equation}

\begin{equation}
  \mathrm{O_2} \rightarrow \mathrm{O}_2^+ + e^-
\end{equation}

\begin{equation}
  \mathrm{O}_3 \rightarrow \mathrm{O}_3^+ + e^-.
\end{equation}

For ionization, the energy deducted from the colliding ion is equal to the sum
of the ionization threshold, plus an amount equal to the energy of the
resulting secondary electron, which is selected randomly from a skewed Gaussian
probability density function\citep{pimblottlavernemozumder} with a mean value
of 33 eV, the average energy per ion pair for O$_2$ \citep{dalgarno1962}.

Shown in Fig. \ref{f1} are the cross-sections as a function of proton energy
for collisions with the neutral species we consider in this work, O, O$_2$, and
O$_3$ using the parameters listed in Tables \ref{proton_ionization_parameters}
and \ref{proton_excitation_parameters}. For atomic and molecular oxygen, we
have used data from \citet{edgar1975}. Due to the lack of proton-ozone
collisional data, in this work, we have fit measured electron impact ionization
\citep{newson1995} and electron impact excitation cross-sections
\citep{sweeney1996} to Eq.  \eqref{green-mcneal-ion} for use in both
proton-ozone and secondary electron-ozone ionization and excitation
collisions.  As can be seen in Fig. \ref{f1}, for all of the neutral species
considered, the ionization cross-sections of protons for all energies above 100
eV are at least about an order of magnitude larger than either the excitation
or elastic collisional cross-sections. Below this energy, elastic collisions
are dominant.  Fig. \ref{f1} also shows the well-known inverse relationship
between primary ion energy and total collisional cross-section, i.e. that
colliding particles with more energy interact less as they travel through
materials than do those with lower energies. This illustrates partially why
attempts to measure the galactic cosmic ray energy spectrum are thwarted by the
effects of Sun, since the lower energy cosmic rays interact more strongly with
the Solar wind \citep{parker1958}. 

\subsubsection{Secondary Electrons}

Once formed, secondary electrons are placed on a random lattice site
next to their parent cations and diffuse away until they are stopped by the
solid through energy lost in inelastic collisions. To calculate their
trajectories in the material, we utilize a hybrid approach which combines
elements of proton travel with our treatment of the hopping of neutral species.
As with protons, we calculate a mean-free-path between inelastic collisions;
however, in the case of electrons, we use this distance to determine the number
of instantaneous jumps from one neighboring lattice site to the next until
ionization or excitation occurs. The value used in determining these distances
is the sum of the total inelastic collisional cross-section and a constant
elastic cross-section equal to the geometrical hard-sphere value of
$\sigma^e_\mathrm{elastic}=10^{-16}$ cm$^2$. The actual distance a secondary
electron travels between collisions, $\Delta x$, is then calculated as in Eq.
\eqref{track.eps}. Energy loss due to elastic collisions of electrons is not
treated rigorously since, for the purposes of chemical changes in the solids,
such collisions are of comparatively less importance than excitations and
ionizations, due to the mass difference between electrons and bulk species.

The secondary electrons formed by the primary ion are known as
``first-generation'' secondary electrons and these can, in turn, ionize other
species in the bulk, resulting in the formation of later generations of
secondary electrons. In this way, a cascade of a few to up to $\sim10^4$
ion-pairs can be formed from a single primary ion \citep{mason2014}. For
electron-impact ionization cross sections, we use the semi-empirical formala
described in \citet{green1972} given by:

\begin{widetext}
  \begin{equation}
    \sigma_\mathrm{i}(E) = \sigma_0 \, A(E)\,\Gamma(E)\,\left[ \mathrm{arctan}\left(\frac{T_\mathrm{m}(E)-T_0(E)}{\Gamma(E)}\right) +  \mathrm{arctan}\left(\frac{T_0(E)}{\Gamma(E)}\right) \right]
    \label{greensawada}
  \end{equation}
\end{widetext}

\noindent 
where, $\sigma_0$ is $10^{-16}$ cm$^{2}$ and $A(E)$ is the differential
cross-section with respect to energy, calculated using 

\begin{equation}
  A(E) = \left( \frac{K}{E} + K_\mathrm{B} \right)\mathrm{ln}\left(\frac{E}{J}+J_\mathrm{B}+\frac{J_\mathrm{C}}{E}\right).
\end{equation}

\noindent
The values $\Gamma(E)$ and $T_0(E)$ are what \citet{green1972} refer to as
width and resonance factors, respectively, and are given by

\begin{equation}
  \Gamma(E) = \Gamma_\mathrm{S}\frac{E}{E+\Gamma_\mathrm{B}}
\end{equation}

\begin{equation}
  T_0(E) = T_\mathrm{S} - \left(\frac{T_\mathrm{A}}{E+T_\mathrm{B}}\right)\mathrm{eV}.
\end{equation}

\noindent
We assume the more energetic electron, post collision, to be the particle which
caused the ionization. Thus, $T_\mathrm{m}$, the maximum energy of the new secondary
electron, is half the remaining energy of the parent electron after the
ionization, i.e.  $T_\textrm{m}=1/2(E-I)$,  where $E$ the energy of the
colliding secondary electron and $I$ the ionization threshold. Values used in
calculating these cross-sections are given in Table
\ref{electron_ionization_parameters}.

  \begin{table*}[ht!]
    \small
    \caption{Electron ionization cross-section parameters}
    \label{electron_ionization_parameters} 
    \begin{tabular*}{\textwidth}{@{\extracolsep{\fill}}ccccccccccc}
      \hline
      State &  $K$ (eV) & $K_\mathrm{B}$ & $J$ (eV) & $J_\mathrm{B}$ & $J_\mathrm{C}$ (eV) & $\Gamma_\mathrm{S}$ (eV) & $\Gamma_\mathrm{B}$ (eV) & $T_\mathrm{S}$ (eV) & $T_\mathrm{A}$ (eV) & $T_\mathrm{B}$ (eV) \\
      \hline
      \multicolumn{11}{c}{Molecular Oxygen\footnotemark[4]} \\
      $X\;^2\Pi_\mathrm{g}$ & 0.48 & 0.00 & 3.76 & 0.00 & 0.00 & 18.50 & 12.10 & 1.86 & 1000.00 & 24.20 \\
      \hline
      \multicolumn{11}{c}{Atomic Oxygen\footnotemark[4]} \\
      $^3\mathrm{P}_2$ & 1.03 & 0.00 & 1.81 & 0.00 & 0.00 & 13.00 & -0.81$_5$ & 6.41 & 3450.00 & 162.00 \\
      \hline 
    \end{tabular*}
  \end{table*}

For electron-impact excitations of bulk species from the ground state to the
$j$-th state, we use the formalism developed by \citet{porter1976} for allowed
transitions, in which the cross-section is given by

\begin{equation}
  \sigma_\mathrm{j}(E) = \left(\frac{q_0\,F_\mathrm{j}\left[1-\left(\frac{W}{E}\right)^\alpha\right]^\beta}{E\,W}\right)\mathrm{ln}\left(\frac{4E\,C}{W} + e \right).
  \label{porterjackmanrgeen}
\end{equation}

\noindent
Here $q_0 = 4\pi a^2 R^2$, where is $R$ the Rydberg energy for atomic hydrogen,
$F_\mathrm{j}$ is the optical oscillator strength, and $\alpha$ and $\beta$
dimensionless values which \citet{porter1976} extracted from experiment. $W$ is
again the excitation energy threshold and ensures proper shape in the
low-energy regime while $C$ ensures proper high-energy falloff.
Species-specific values used in this work for calculating these excitation
cross-sections can be found in Tables \ref{electron_allowed_parameters} and
\ref{electron_forbidden_parameters}. 

  \begin{table}[ht]
    \small
    \caption{Electron excitation cross-section parameters for allowed transitions}
    \label{electron_allowed_parameters}
    \begin{tabular*}{0.5\textwidth}{@{\extracolsep{\fill}}cccccc}
      \hline
      Excited State & $W$(eV) & $\alpha$ & $\beta$ & $C$ & $F$ \\
      \hline
      \multicolumn{6}{c}{Atomic Oxygen\footnotemark[4]} \\
      $^3\mathrm{S}$ & 9.53 & 0.86 & 1.44 &  0.32 & 0.056 \\
      \hline
      \multicolumn{6}{c}{Molecular Oxygen\footnotemark[5]} \\
      $B^3\Sigma_\mathrm{u}^-$ & 8.4 & 1.19 & 2.31 & 0.037 & 0.25 \\
      9.9 eV peak & 9.9 & 1.38 & 3.44 & 0.62 & 0.029 \\
      \hline
    \end{tabular*}
  \end{table}

\footnotetext[4]{Data taken from \citet{jackman1977}}
\footnotetext[5]{Data taken from \citet{porter1976}}

  \begin{table}[htb]
    \small
    \caption{Electron excitation cross-section parameters for forbidden transitions}
    \label{electron_forbidden_parameters} 
    \begin{tabular*}{0.5\textwidth}{@{\extracolsep{\fill}}cccccc}
      \hline
      Excited State & $W$(eV) & $\alpha$ & $\beta$ & $\Omega$ & $F$ \\
      \hline
      \multicolumn{6}{c}{Atomic Oxygen\footnotemark[4]} \\
      $^1\mathrm{D}$ & 1.96 & 1.00 & 2.00 & 1.00 & 0.010 \\
      $^1\mathrm{S}$ & 4.18 & 0.50 & 1.00 & 1.00 & 0.0042 \\
      $^5\mathrm{S}$ & 10.60 & 19.20 & 10.50 & 2.69 & 0.013 \\
      \hline
      \multicolumn{6}{c}{Molecular Oxygen\footnotemark[5]} \\
      $a^1\Delta_\mathrm{g}$ & 0.98 & 3.00 & 1.00 & 3.00 & 0.0005 \\
      $b^1\Sigma_\mathrm{g}^+$ & 1.64 & 3.00 & 1.00 & 3.00 & 0.0005 \\
      $A^3\Sigma_\mathrm{u}^+$ & 4.50 & 1.00 & 1.00 & 0.90 & 0.021 \\
      \hline  
    \end{tabular*}
  \end{table}

\begin{figure}[htb!]
  \centering
  \begin{subfigure}[t]{0.48\textwidth}
    \includegraphics[height=5.5cm]{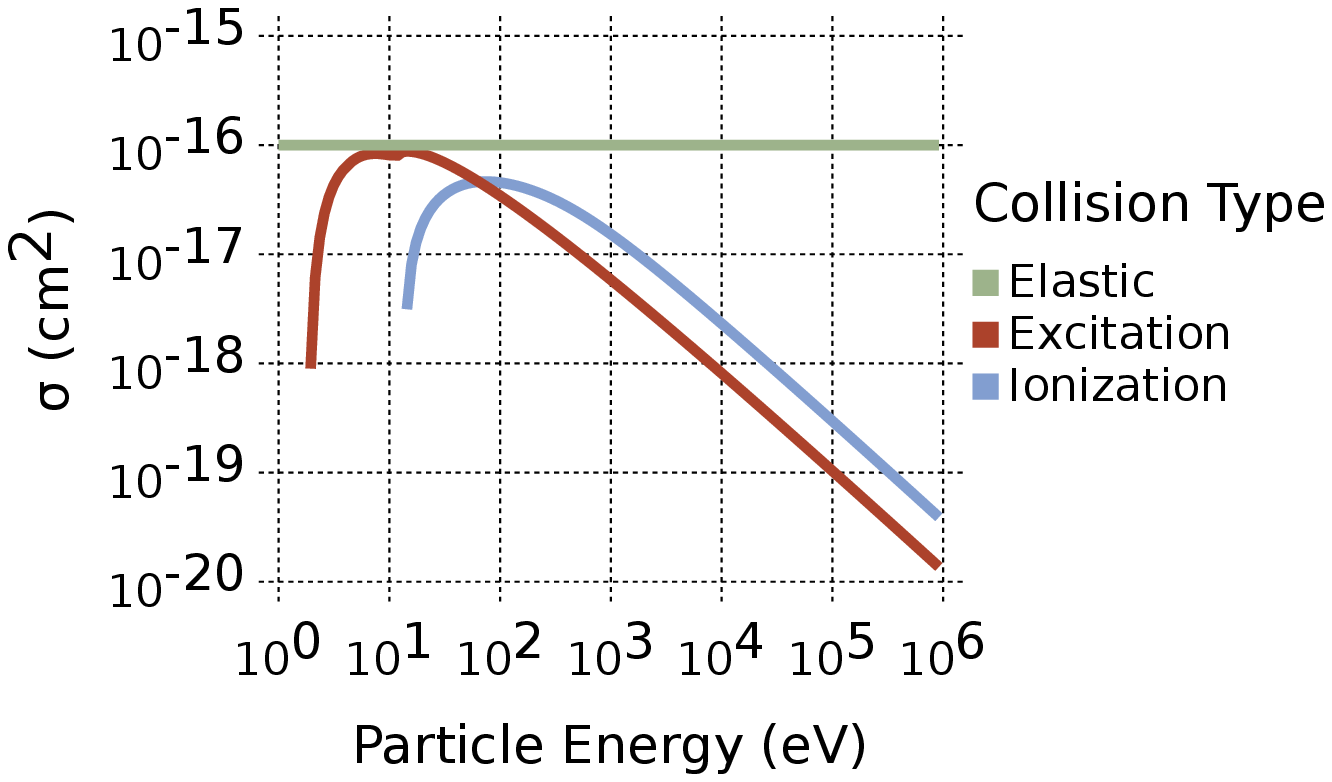}
    \caption{Atomic oxygen}
    \label{f2a}
  \end{subfigure}
  \begin{subfigure}[t]{0.48\textwidth}
    \includegraphics[height=5.5cm]{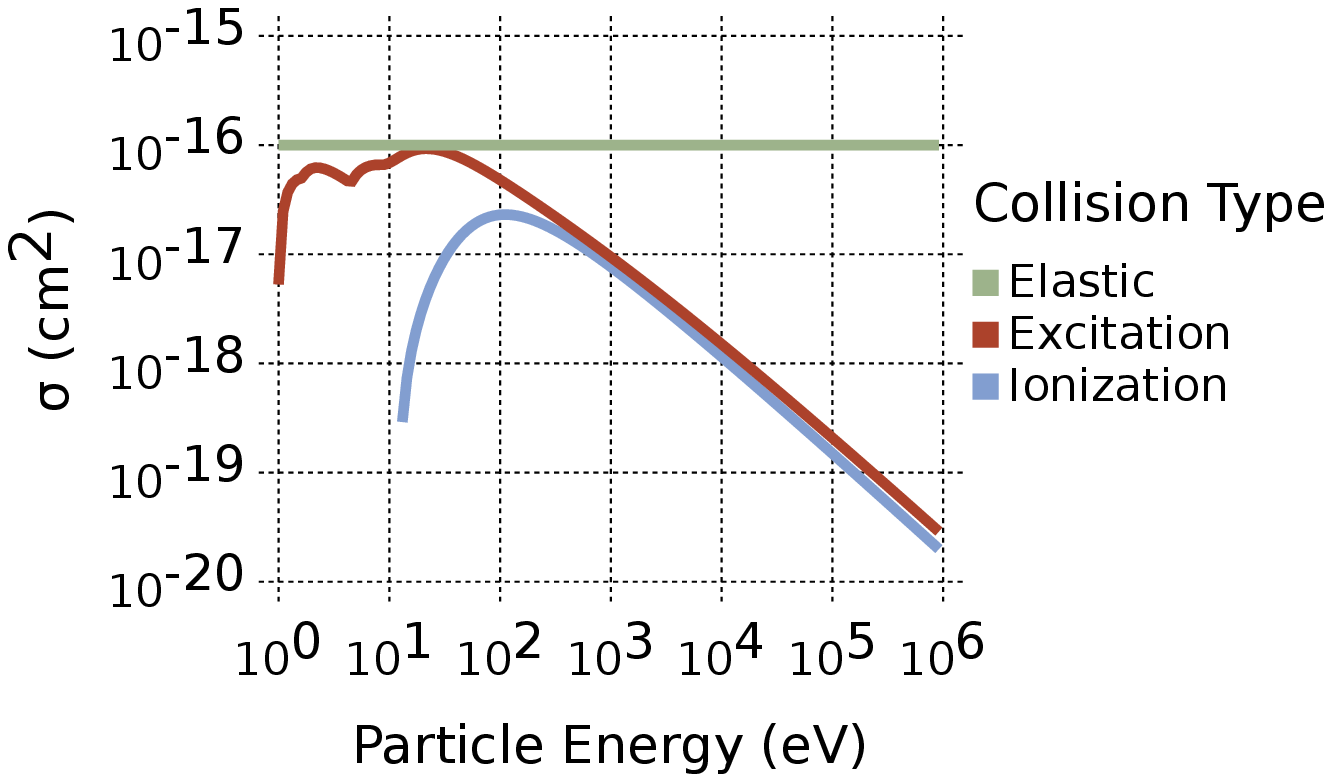}
    \caption{Molecular oxygen}
    \label{f2b}
  \end{subfigure}
  \begin{subfigure}[t]{0.48\textwidth}
    \includegraphics[height=5.5cm]{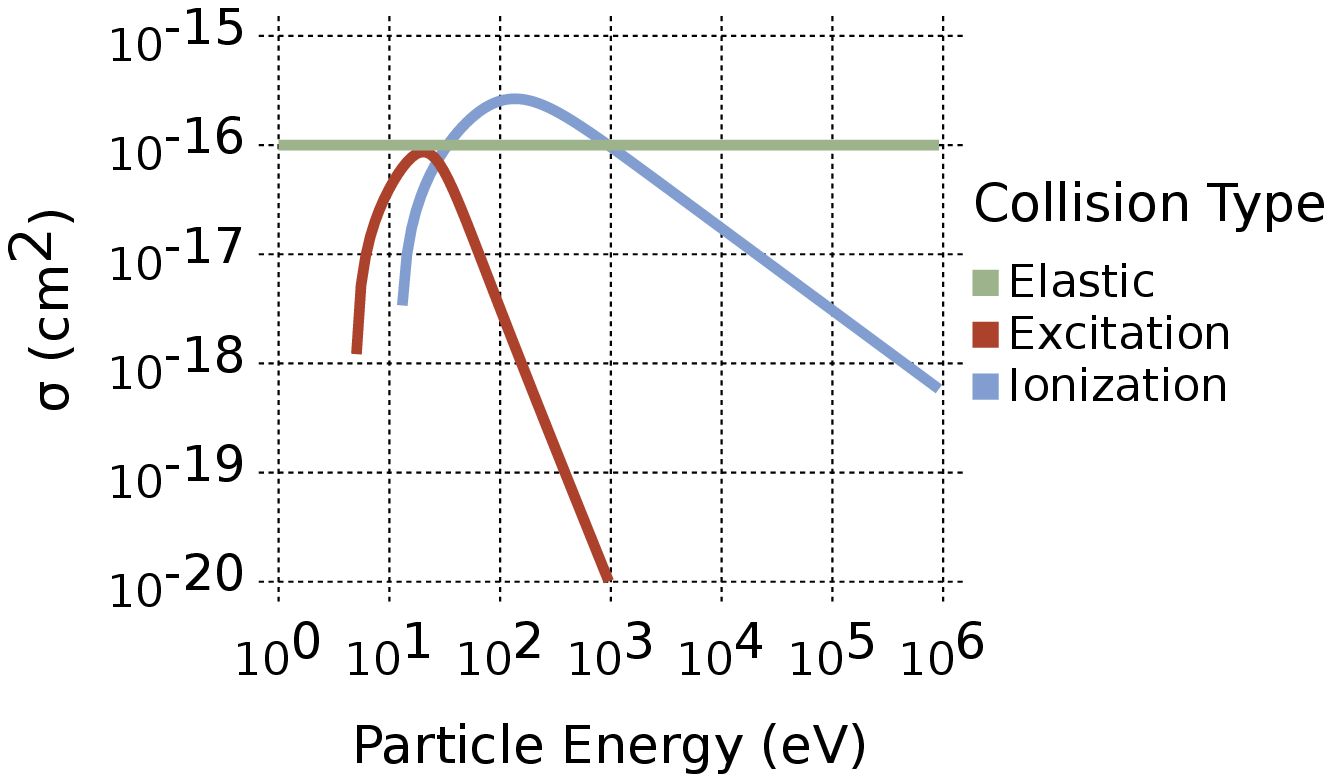}
    \caption{Ozone}
    \label{f2c}
  \end{subfigure}
  \caption{Electron-collision cross-sections calculated using parameters from Tables \ref{electron_ionization_parameters}, \ref{electron_allowed_parameters}, and \ref{electron_forbidden_parameters}.}
  \label{f2}
\end{figure}

Our calculated values for the electron cross-sections, using the parameters
given in Tables \ref{electron_ionization_parameters},
\ref{electron_allowed_parameters}, and \ref{electron_forbidden_parameters}, are
shown in Fig. \ref{f2}. Comparison with Fig. \ref{f1} shows that the
cross-sections for electron impact are generally smaller than the equivalent
values for protons at the same energy. 

Energy is deducted from secondary electrons until a lower threshold of 4 eV is
reached, at which point a secondary electron enters the sub-excitation regime.
Sub-excitation secondary electrons are thought to be the delivery mechanism for
$\sim$10-20\% of the energy deposited in the solid by the primary ion
\citep{fano1986}, and affect the chemistry through a variety of mechanisms,
such as the resonant process of dissociative electron attachment (DEA)
\citep{arumainayagam2010}. We choose 4 eV as our lower energy threshold since
it corresponds approximately to the low energy limit of the resonance which
leads to DEA in O$_2$ \citep{arumainayagam2010}.

As listed in Table \ref{nonthermal}, we assume that sub-excitation electrons
are destroyed in one of two ways.  One possibility is through charge
recombination with a neighboring cation.  The other is by reaction with a
randomly chosen neighboring neutral species via DEA to form an anion. These
anions then react with nearby cations formed in previous ionization collisions
at the end of the track calculation phase, at which time all remaining charged
species are neutralized by such fast ion-ion reactions \citep{johnson1990}.

  \begin{table}[htb!]
    \small
    \caption{Secondary electron reactions}
    \label{nonthermal}
    \begin{tabular*}{0.48\textwidth}{@{\extracolsep{\fill}}llll}
      \hline
      ID\# & Electron Attachment\footnotemark[6] & $f_\mathrm{branching}$ & Source \\
      \hline
      1 & $\mathrm{O} + e^- \rightarrow \mathrm{O}^- $ & 1.00 & \citep{phelps1969} \\
      2 & $\mathrm{O}_2 + e^- \rightarrow \mathrm{O}^- + \mathrm{O}$ & 1.00 & \citep{arumainayagam2010} \\
      3 & $\mathrm{O}_3 + e^- \rightarrow \mathrm{O}^- + \mathrm{O_2}$ & 1.00 & \citep{senn1999} \\
      \hline
      ID\# & Electron Recombination & $f_\mathrm{branching}$ & Source  \\

      \hline
      4 & $\mathrm{O}^+ + e^- \rightarrow \mathrm{O(^1\mathrm{D})} $ & 1.00 & \citep{garrod2008}  \\
      5 & $\mathrm{O}_2^+ + e^- \rightarrow 2\mathrm{O}$ & 0.20 & \citep{peverall2001}  \\
      6 & $\mathrm{O}_2^+ + e^- \rightarrow 2\mathrm{O(^1\mathrm{D})}$ & 0.36 & \citep{peverall2001}  \\
      7 & $\mathrm{O}_2^+ + e^- \rightarrow \mathrm{O} + \mathrm{O(^1\mathrm{D})}$ & 0.44 & \citep{peverall2001}  \\
      8 & $\mathrm{O}_3^+ + e^- \rightarrow \mathrm{O_2} + \mathrm{O}$ & 0.06 & \citep{zhaunerchyk2008} \\
      9 & $\mathrm{O}_3^+ + e^- \rightarrow 3\mathrm{O}$ & 0.26 & \citep{zhaunerchyk2008}  \\
      10 & $\mathrm{O}_3^+ + e^- \rightarrow 3\mathrm{O(^1\mathrm{D})}$ & 0.0094 & \citep{zhaunerchyk2008}  \\
      11 & $\mathrm{O}_3^+ + e^- \rightarrow \mathrm{O} + 2\mathrm{O(^1\mathrm{D})}$ & 0.20 & \citep{zhaunerchyk2008}  \\
      12 & $\mathrm{O}_3^+ + e^- \rightarrow 2\mathrm{O} + \mathrm{O(^1\mathrm{D})}$ & 0.47 & \citep{zhaunerchyk2008}  \\
      \hline
    \end{tabular*}
  \end{table}

\subsection{Continuous-time random walk chemical modeling} \label{chem-phase}

In the chemical phase of the model, ground-state neutral species in the bulk
can diffuse throughout the crystal lattice of the solid and reactions can
occur. The Monte Carlo technique used for modeling the chemistry of the solid
is the continuous-time random walk method initially developed by
\citet{montroll1965}, which was further developed by \citet{chang2005} and
others \citep{chang2014, chang2016, hincelin2015, lauck2015} for studying
molecular formation on interstellar dust grains. In this scheme, the
interstitial sites facilitate bulk diffusion by allowing species to hop from
one interstitial site to another with a low barrier. Normal lattice species
are, effectively, immobile and do not become interstitial species; however, if
a species on a normal lattice site dissociates or reacts, the products can
become interstitial.  Moreover, an interstitial species can hop to a lattice
defect site to become a normal lattice species. A cartoon representation of
these possible moves is given in Fig. 1 of \citet{chang2014}.

Hopping rates are parameterized by three species-specific values: the
binding energy for surface species, $E_\mathrm{D}$; the barrier against
diffusion for surface species, $E_\mathrm{b1}$; and, the barrier against
diffusion for bulk species, $E_\mathrm{b2}$. We approximate the surface barrier
against diffusion as being $E_\mathrm{b1} = 0.5E_\mathrm{D}$ following the
approximation used in \citet{garrod2006}. The bulk barrier against diffusion is
assumed to be $E_\mathrm{b2} = 0.7E_\mathrm{D}$ \citep{chang2014}.

For any surface species, the rate coefficient for hopping, $b_1$, given by

\begin{equation}
  b_1 = \nu\;\mathrm{exp}\left( -\frac{E_\mathrm{b1}}{T} \right)
  \label{b1}
\end{equation}

\noindent 
where $T$ is the temperature of the solid and $\nu$ is the trial frequency,
which has a value of \trialfrequency \citep{chang2014, hasegawa1992}.
Similarly, the rate coefficient for bulk species is

\begin{equation}
  b_2 = \nu \; \mathrm{exp}\left( -\frac{E_\mathrm{b2}}{T} \right).
  \label{b2}
\end{equation}

\noindent 
Moreover, on the surface, our Monte Carlo technique allows for the possibility
of desorption into the gas phase with a rate coefficient of

\begin{equation}
  b_3 = \nu \; \mathrm{exp}\left( -\frac{E_\mathrm{D}}{T} \right).
  \label{b3}
\end{equation}

\noindent 
Whether a surface species hops or desorbs is based on a competitive mechanism
based on a random number, $R_\mathrm{D}$, to determine the outcome:

  \begin{eqnarray} \label{competition}
    \text{Outcome}
    \begin{cases}
      & \text{Desorption for } 0 < R_\mathrm{D} \leq \frac{b_3}{b_3 + b_1} \\
      & \text{Diffusion for } \frac{b_3}{b_3 + b_1} < R_\mathrm{D} \leq 1
    \end{cases}
  \end{eqnarray}

\noindent
Because the primary focus of this work is on modeling the 
chemistry of the irradiated O$_2$ system, and because the model is not coupled to any gas-phase
chemical system, we have disabled the desorption of surface species. Even when enabled, however,
such thermal desorption is negligible, given the low ice temperature of 5 K and the 
high reactivity of the most weakly bound species, atomic oxygen.

\footnotetext[6]{Only applicable for sub-excitation electrons}

Reactions in the model are assumed to proceed via the diffusive
(Langmuir-Hinshelwood) mechanism, and can occur when species hop to sites
occupied by possible co-reactants, and products can be formed. If the reaction
has a barrier, a competitive mechanism is also utilized to determine whether
the reaction proceeds or the hopping species diffuses away.  If one of the
co-reactants occupies a normal bulk lattice site and there are multiple
products, the one with the higher binding energy will be placed at the site,
with the rest being randomly placed in the surrounding interstitial sites.  On
the surface, only the normal lattice sites can be occupied, i.e. it is assumed
that there are no available interstitial sites on the surface.

The waiting time between hops, $\tau$, for a species on the surface is given by

\begin{equation}
  \tau_\mathrm{surf} = -\frac{\mathrm{ln}(R_\mathrm{D})}{b_1 + b_3},
\end{equation}

\noindent
while for bulk species, the time is

\begin{equation}
  \tau_\mathrm{bulk} = -\frac{\mathrm{ln}(R_\mathrm{D})}{b_2}.
\end{equation}

\begin{figure}[htb]
  \centering
   \includegraphics[height=6.5cm]{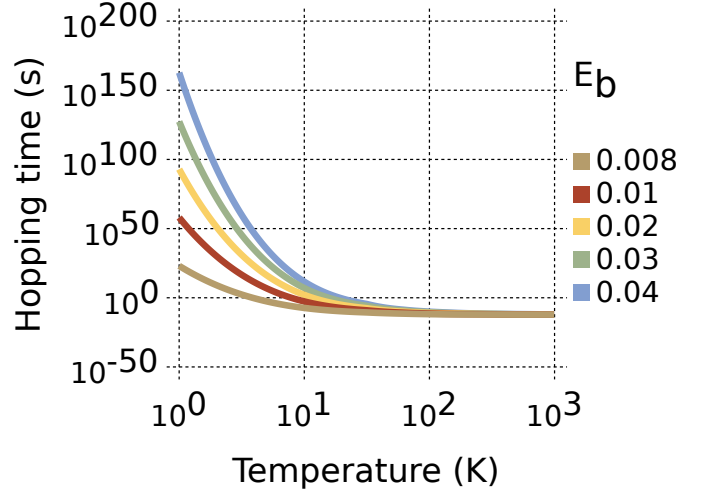}
  \caption{Hopping times for neutral species as a function of system temperature and diffusion barrier height. }
  \label{f3}
\end{figure}

\noindent
As shown in Fig. \ref{f3}, bulk hopping times are affected by both the system
temperature and the height of the diffusion barrier. Above about 100 K, further
increases in temperature have little or no effect on the hopping time; however,
there is a strong temperature dependence below 10 K and this effect becomes
larger with increasing diffusion barriers. As described in detail in
\citet{chang2014}, the waiting time until the next action is stored in a data
structure that also contains the type and current coordinates of each species
in the system. After a hop, the model searches the waiting list and the species
associated with the shortest waiting moves next.

\subsection{Chemical Network} \label{network}

In their work, B99 fit their experimental data with a simplified chemical
network consisting of:

\begin{equation}
  \mathrm{O_2} \rightarrow 2\mathrm{O}
  \label{o2-dis}
\end{equation}

\begin{equation}
  \mathrm{O_3} \rightarrow \mathrm{O} + \mathrm{O_2}
  \label{o3-dis}
\end{equation}

\begin{equation}
  \mathrm{O} + 2\mathrm{O}_2 \rightarrow \mathrm{O}_3 + \mathrm{O_2}
  \label{o-o2}
\end{equation}

\begin{equation}
  \mathrm{O} + \mathrm{O}_3 \rightarrow \mathrm{O}_2 + \mathrm{O}_2
  \label{o-o3}
\end{equation}

\noindent
where here, reactions \eqref{o2-dis} and \eqref{o3-dis} are irradiation induced
dissociations, and reactions \eqref{o-o2} and \eqref{o-o3} are assumed to
proceed via thermal diffusion in the solid. Our network expands on the work of
B99, in part, by including electronically excited neutral species, as shown in
Table \ref{thermal}. We are able to do this since our method of calculating the
tracks of the proton and secondary electrons allows us to determine the final
state of a species after undergoing an excitation collision; however, our
ability to do so is limited by the scarcity of data on the reactivity and
product pathways of species in most of these excited states. For instance,
though we explicitly consider excitation of O$_2$ from its triplet ground state
to both excited singlet and triplet states, as shown by the cross-sectional
parameters in Tables
\ref{proton_excitation_parameters}-\ref{electron_forbidden_parameters}, the
available theoretical and experimental kinetic data is almost entirely for the
$^1\Delta$ state. Given this discrepancy between the available cross-sectional
and kinetic data, in our network, we denote all electronically excited O$_2$ as
O$_2$($^1\Delta$). We similarly denote the other excited neutral species in our
network using the term symbol of the best studied higher state.

  \begin{table}[htb!]
    \small
    \caption{Chemical network. The activation energies and branching fractions are taken from gas-phase studies.}
    \label{thermal}
    \begin{tabular*}{0.495\textwidth}{@{\extracolsep{\fill}}lllll}
      \hline
      ID\# & Neutral-Neutral Reactions & $f_\mathrm{branching}$ & $E_\mathrm{A}$(kJ/mol) & Source \\
      \hline
      13 & $\mathrm{O} + \mathrm{O} \rightarrow \mathrm{O_2(^1\Delta)}$ & 1.00 & 0.00 & \citep{warnatz1984}  \\
      14 & $\mathrm{O} + \mathrm{O_2} \rightarrow \mathrm{O_3(^1\mathrm{B}_2)}$ & 1.00 & 0.00 & \citep{demore1997} \\
      15 & $\mathrm{O} + \mathrm{O_3} \rightarrow 2\mathrm{O}_2$  & 1.00 & 17.12 & \citep{atkinson2004} \\
      16 & $\mathrm{O_3} + \mathrm{O_3} \rightarrow 3\mathrm{O_2}$ & 1.00 & 77.41 & \citep{pshezhetskii1959} \\
      17 & $\mathrm{O(^1D)} + \mathrm{O} \rightarrow 2\mathrm{O}$ & 0.50 & 0.00  & \citep{sobral1993} \\
      18 & $\mathrm{O(^1D)} + \mathrm{O} \rightarrow  \mathrm{O_2(^1\Delta)}$ & 0.50 & 0.00 & \begin{small}See text\end{small} \\
      19 & $\mathrm{O(^1D)} + \mathrm{O_2} \rightarrow \mathrm{O} + \mathrm{O_2}$ & 0.50 & 0.00 & \citep{demore1997} \\
      20 & $\mathrm{O(^1D)} + \mathrm{O_2} \rightarrow \mathrm{O_3(^1\mathrm{B}_2)}$ & 0.50 & 0.00 & \begin{small}See text\end{small} \\
      21 & $\mathrm{O(^1D)} + \mathrm{O_3} \rightarrow 2\mathrm{O}_2$  & 1.00 & 0.00 & \citep{brasseur1984} \\
      22 & $\mathrm{O_2(^1\Delta)} + \mathrm{O} \rightarrow \mathrm{O} + \mathrm{O_2}$ & 0.50 & 0.00 & \citep{doroshenko1992} \\
      23 & $\mathrm{O_2(^1\Delta)} + \mathrm{O} \rightarrow \mathrm{O_3(^1\mathrm{B}_2)}$ & 0.50 & 0.00 & \begin{small}See text\end{small} \\
      24 & $\mathrm{O_2(^1\Delta)} + \mathrm{O_2} \rightarrow 2\mathrm{O_2}$ & 1.00 & 0.00 & \citep{klopovskiy1999} \\
      25 & $\mathrm{O_2(^1\Delta)} + \mathrm{O_3} \rightarrow 2\mathrm{O_2} + \mathrm{O}$ & 1.00 & 23.61 & \citep{demore1997} \\
      26 & $\mathrm{O(^1D)} + \mathrm{O(^1D)} \rightarrow 2\mathrm{O}$ & 0.50 & 0.00 & \begin{small}See text\end{small} \\
      27 & $\mathrm{O(^1D)} + \mathrm{O(^1D)} \rightarrow \mathrm{O_2(^1\Delta)}$ & 0.50 & 0.00 & \begin{small}See text\end{small} \\
      28 & $\mathrm{O(^1D)} + \mathrm{O_2(^1\Delta)} \rightarrow \mathrm{O} + \mathrm{O_2}$ & 0.50 & 0.00 & \citep{doroshenko1992} \\
      29 & $\mathrm{O(^1D)} + \mathrm{O_2(^1\Delta)} \rightarrow \mathrm{O_3(^1\mathrm{B}_2)}$ & 0.50 & 0.00 & \begin{small}See text\end{small} \\
      30 & $\mathrm{O(^1D)} + \mathrm{O_3(^1\mathrm{B}_2)} \rightarrow 2\mathrm{O}_2$  & 1.00 & 0.00 & \begin{small}See text\end{small} \\
      31 & $\mathrm{O_2(^1\Delta)} + \mathrm{O_2(^1\Delta)} \rightarrow 2\mathrm{O_2}$ & 1.00 & 3.23 & \citep{heidner1981} \\
      32 & $\mathrm{O_2(^1\Delta)} + \mathrm{O_3(^1\mathrm{B}_2)} \rightarrow 2\mathrm{O_2} + \mathrm{O}$ & 1.00 & 23.61 & \begin{small}See text\end{small} \\
      33 & $\mathrm{O_3(^1\mathrm{B}_2)} + \mathrm{O_3(^1\mathrm{B}_2)} \rightarrow 3\mathrm{O_2}$ & 1.00 &   77.41 & \begin{small}See text\end{small} \\
      34 & $\mathrm{O_3(^1\mathrm{B}_2)} + \mathrm{O} \rightarrow 2\mathrm{O}_2$  & 1.00 & $17.12$ & \begin{small}See text\end{small} \\
      35 & $\mathrm{O_3(^1\mathrm{B}_2)} + \mathrm{O_2} \rightarrow 2\mathrm{O}_2 + \mathrm{O}$  & 1.00 & $23.61$ & \begin{small}See text\end{small} \\
      36 & $\mathrm{O_3(^1\mathrm{B}_2)} + \mathrm{O_3} \rightarrow 3\mathrm{O_2}$ & 1.00 & 77.41 & \begin{small}See text\end{small} \\
      \hline
      ID\# & Ion Recombinations & $f_\mathrm{branching}$ & $E_\mathrm{A}$(kJ/mol) & Source \\      \hline
      37 & $\mathrm{O}_2^+ + \mathrm{O}_2^- \rightarrow \mathrm{O_2}(^1\Delta) + \mathrm{O_2}$ & 1.00 & 0.00 &\citep{fridman2008} \\
      38 & $\mathrm{O}_2^+ + \mathrm{O}^- \rightarrow \mathrm{O_2}(^1\Delta) + \mathrm{O}$ & 1.00 & 0.00 & \citep{fueki1963} \\
      39 & $\mathrm{O}_2^- + \mathrm{O}^+ \rightarrow \mathrm{O_2}(^1\Delta) + \mathrm{O}$ & 1.00 & 0.00 & \citep{fueki1963} \\
      40 & $\mathrm{O}^+ + \mathrm{O}^- \rightarrow 2\mathrm{O(^1D)}$ & 1.00 & 0.00 & \begin{small}See text\end{small}\\
      41 & $\mathrm{O^+} + \mathrm{O_3^-} \rightarrow \mathrm{O(^1D)} + \mathrm{O_3(^1\mathrm{B}_2)}$ & 1.00 & 0.00 & \begin{small}See text\end{small}\\
      42 & $\mathrm{O^+} + \mathrm{O_3^-} \rightarrow 2\mathrm{O_2(^1\Delta)}$ & 1.00 & 0.00 & \begin{small}See text\end{small}\\
      43 & $\mathrm{O^-} + \mathrm{O_3^+} \rightarrow \mathrm{O(^1D)} + \mathrm{O_3(^1\mathrm{B}_2)}$ & 1.00 & 0.00 & \begin{small}See text\end{small}\\
      44 & $\mathrm{O^-} + \mathrm{O_3^+} \rightarrow 2\mathrm{O_2(^1\Delta)}$ & 1.00 & 0.00 & \begin{small}See text\end{small}\\
      45 & $\mathrm{O_2^-} + \mathrm{O_3^+} \rightarrow \mathrm{O_3(^1\mathrm{B}_2)} + \mathrm{O_2}(1\Delta)$ & 1.00 & 0.00 & \begin{small}See text\end{small}\\
      46 & $\mathrm{O_2^+} + \mathrm{O_3^-} \rightarrow \mathrm{O_3(^1\mathrm{B}_2)} + \mathrm{O_2}(1\Delta)$ & 1.00 & 0.00 & \begin{small}See text\end{small}\\
      47 & $\mathrm{O_3^+} + \mathrm{O_3^-} \rightarrow 3\mathrm{O_2(^1\Delta)}$ & 1.00 & 0.00 & \begin{small}See text\end{small}\\
      \hline

    \end{tabular*}
  \end{table}

In our code, we assume that neutrals in excited electronic states either react
immediately with a neighboring species or relax back to the ground state if
there are no possible nearest neighbor co-reactants. The important role excited
species play in the chemistry of the irradiated oxygen ice system is
illustrated by the following two reactions:

$$
  \mathrm{O}(^3\mathrm{P}) + \mathrm{O_3} \rightarrow 2\mathrm{O_2} 
$$

$$
  \mathrm{O}(^1\mathrm{D}) + \mathrm{O_3} \rightarrow 2\mathrm{O_2}.
$$

\noindent
Here, the reaction between ozone and the triplet ground state of atomic oxygen
has been measured to have a barrier of about 0.2 eV \citep{atkinson2004},
whereas no barrier has been reported for the reaction between excited singlet
atomic oxygen and ozone \citep{demore1997}.

Following B99, we have drawn heavily on previous research in atmospheric
chemistry in compiling our network
\citep{brasseur1984,demore1997,atkinson2004}; however, in some cases we have
been unable to find data on certain reactions, particularly for those involving
two electronically excited species. For these, we have based the product
channels and branching fractions on similar reactions. Since the
chemistry here is occurring in an ice, we have added additional pathways to some
of the gas-phase reactions to account for solid-phase effects.  For
instance, to the previously studied \citep{sobral1993} gas-phase reaction

\begin{equation*}
  \mathrm{O} + \mathrm{O}(^1\mathrm{D}) \rightarrow 2\mathrm{O}
\end{equation*}

\noindent
we have added the additional product channel

\begin{equation*}
  \mathrm{O} + \mathrm{O}(^1\mathrm{D}) \rightarrow \mathrm{O_2}
\end{equation*}

\noindent
to account for the trapping and stabilization of products in the ice.

Ion-ion recombinations present unique challenges in our model, since there are
a number of product channels involving species that can be formed in a number
of different excited states. For simplicity, where we have been unable to find
data on these processes, we base the product channels on the corresponding
neutral-neutral reactions, but with products assumed to be electronically
excited, as with:

\begin{equation*}
  \mathrm{O^-} + \mathrm{O_3^+} \rightarrow 2\mathrm{O_2(^1\Delta)}
\end{equation*}

\begin{equation*}
  \mathrm{O^+} + \mathrm{O_3^-} \rightarrow 2\mathrm{O_2(^1\Delta)},
\end{equation*}

\noindent
which were based on the reaction between singlet atomic oxygen and
ozone. 

One class of reactions not present in our network consists of those between
ions and neutral species. These are often barrierless and are known to be of
significant astrochemical importance \citep{wakelam2010}. We do not include
these since we assume in this work that ions are quickly neutralized via charge
recombination reactions, as suggested by previous theoretical and experimental
studies \citep{baragiola1999,johnson1990}.  This is done due to the uncertain
diffusion barriers of charged species in solids, and the significant influence
of effects which we do not consider in this work, such as Coulombic forces.  

\section{Results and Discussion}

As an initial test of our approach, we simulated the experiment
described in B99 in which O$_2$ is partially converted to O$_3$.
Listed in Table \ref{parameters} are the physical conditions we take from that
work. We assume an initial pure O$_2$ ice at a constant temperature of
5 K, connected to a helium cryostat. The ice studied in B99 had a thickness of
$\sim10\;\mu m$. In this work we model a $\sim0.1\;\mu$m thick ice, due both to
computational expense and because our interest is ultimately in the chemistry
of the ice mantles of interstellar dust grains, which are thought to reach a
maximum thickness on the order of $\sim 0.01\;\mu$m \citep{herbst2014}.  In our
code, the simulation begins when the first proton collides with the pristine
ice.  Following B99, we assume an initial proton energy of \protonenergy keV
and a flux of \flux. The model continues to follow subsequent proton arrivals
and hopping of species in the ice until an upper fluence limit of $10^{17}$
protons/cm$^{2}$ is reached, where fluence is defined as the total irradiation
exposure per unit surface area.  It is used in comparing the results of
irradiation in a way that is less dependent on the specific flux of particles
to which the material is exposed.

  \begin{table}[htb]
    \small
    \caption{Model parameters}
    \label{parameters}
    \begin{tabular*}{0.48\textwidth}{@{\extracolsep{\fill}}ll}
      \hline
      Physical Conditions from B99 & Value  \\
      \hline
      Temperature & 5 K  \\
      Ice Density [O$_2$] & \rhoIce \\
      Initial Proton Energy & \protonenergy keV \\
      Proton Flux & \protonflux \\
      \hline
      Species & Binding Energy to O$_2$ (eV) \\
      \hline
      O & $2.15\times 10^{-2}$ \\
      O$_2$ & $7.85\times 10^{-2}$ \\
      O$_3$ & $6.41\times 10^{-2}$ \\
      \hline
    \end{tabular*}
  \end{table}

Using the parameters listed in Table \ref{parameters}, our code predicts that
each proton undergoes a collision in the bulk of the ice approximately every
300 \AA, or about three times, given the thickness of the simulated ices in
this work. On average, protons in these simulations lose on the order of
$\sim10^2$ eV, or 0.1$\%$ of their initial energy, by the time they exit the
system.  We can, however, arbitrarily fix the distance between proton
collisions to be some smaller value. By thus increasing the number of
collisions, and thus the energy deposited in the system by each proton, we can
approximate the effects of having a thicker ice. From the formula for the
stopping power, given in Eq. \eqref{stopping}, one can derive the energy lost
by a proton travelling in a straight path through any given solid to be

\begin{equation}
  \Delta E = \Delta x \sum_n \frac{Q_{n}}{\Lambda_{n}},
\end{equation}

\noindent
where $\Delta x$ is some path length and the sum is over $n$ inelastic
processes, with each resulting in an average energy loss, $Q_{n}$, by the
primary ion and  having a mean-free-path of $\Lambda_{n}$. If $\Delta x$ is
the total thickness of the solid, $\Delta E$ approximates the total energy lost
by the primary ion, which is proportional to ice thickness and inversely
proportional to the mean-free-path between energy loss events.

\begin{figure}[htb!]
  \begin{subfigure}[l]{0.48\textwidth}
    \includegraphics[height=5.5cm]{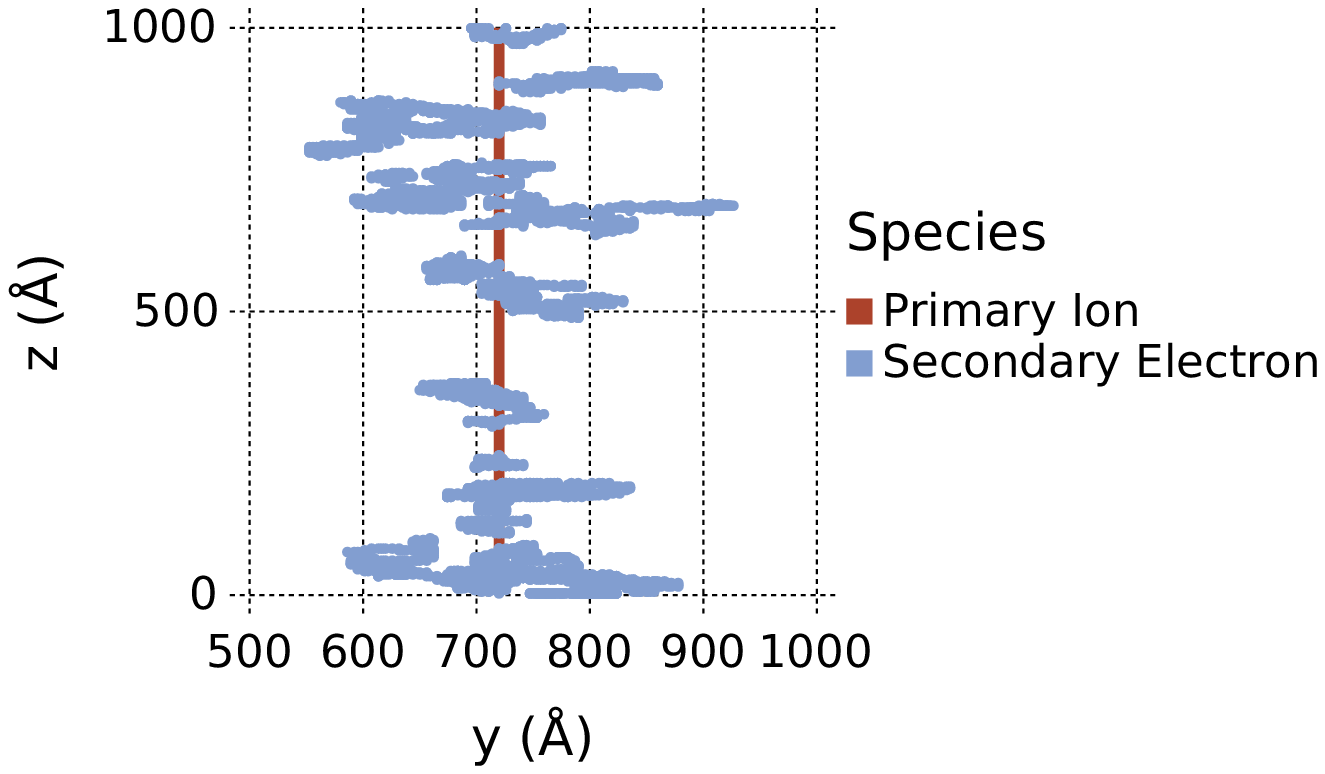}
    \caption{$y$-$z$ track structure}
    \label{f4a}
  \end{subfigure}
  \begin{subfigure}[l]{0.48\textwidth}
    \includegraphics[height=5.5cm]{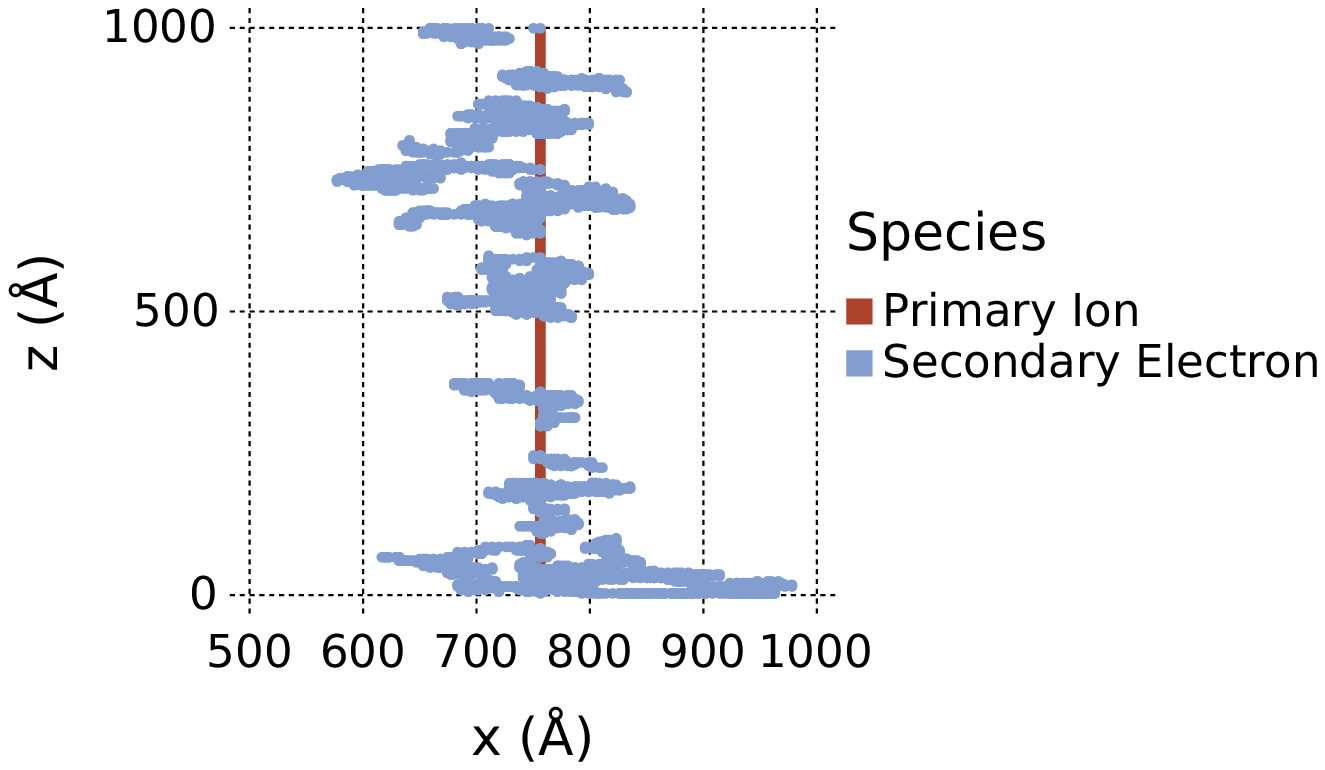}
    \caption{$x$-$z$ track structure}
    \label{f4b}
  \end{subfigure}
  \begin{subfigure}[l]{0.48\textwidth}
    \includegraphics[height=5.5cm]{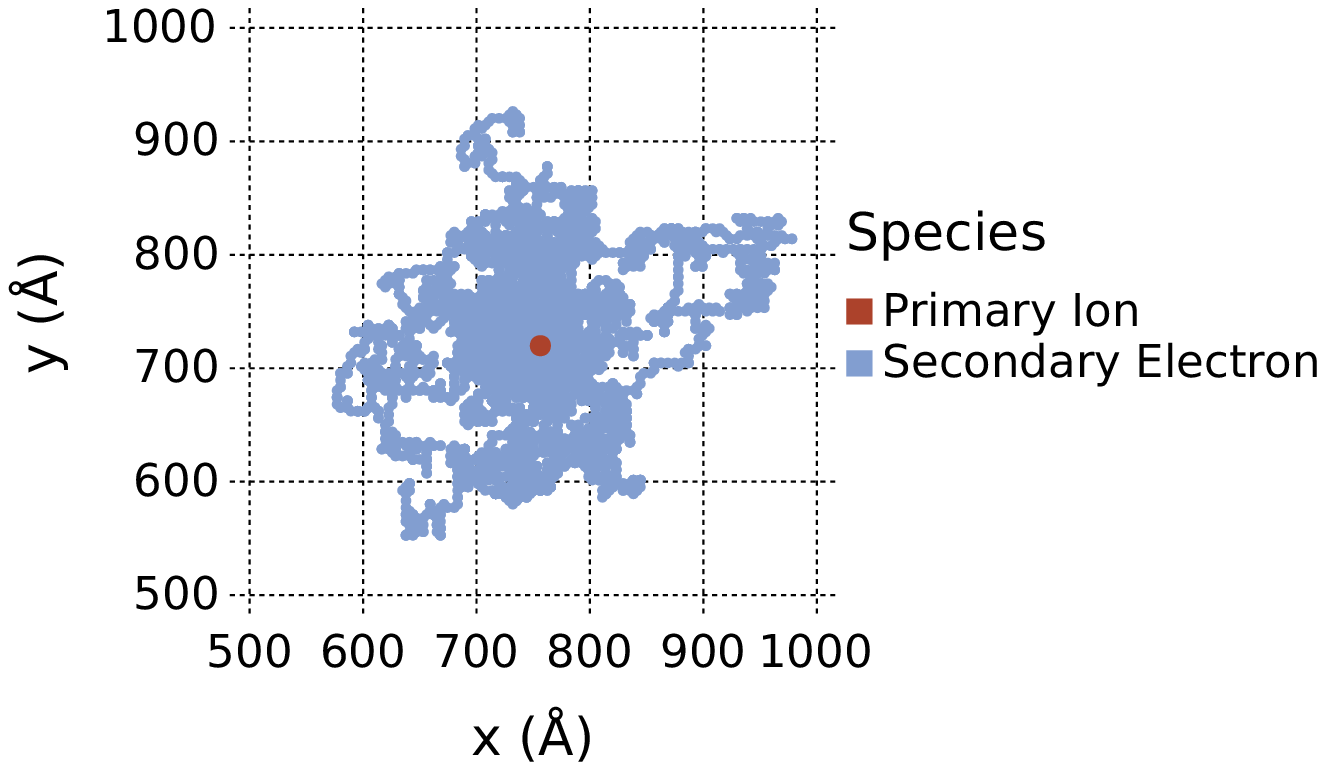}
    \caption{$x$-$y$ track structure}
    \label{f4c}
  \end{subfigure}
  \caption{Sample track structure in which a proton collides with a pristine O$_2$ ice with fixed collision distance of $\Delta x= 33$ \AA.}
  \label{f4}
\end{figure}

Shown in Figs. \ref{f4a}-\ref{f4c} is an example of a track calculation in
which, starting with pristine O$_2$ ice, we fixed the distance between
collisions to be 33 \AA, corresponding to a total cross-section larger than the
real value by about an order of magnitude. Here the proton enters the ice at
$z=0$ \AA, with an entry angle of 90$^\circ$ relative to the surface, and
travels in a nearly straight path through the material until it exits at
$z=1000$ \AA.  In Figs.  \ref{f4a}-\ref{f4c} the path of the proton is
represented by a red line and secondary electron paths are given in blue.
\texttt{CIRIS} performs such a track calculation for every proton arrival.
Given the simulated size of our ice, to reach a fluence of $10^{17}$
protons/cm$^{2}$, our code calculates more than $10^6$ such proton arrivals.

Between track calculations, neutral species diffuse through the solid and can
react with their neighbors. The rate of this diffusive chemistry is governed,
in part, by the hopping rates of each of the reactants.  In our code, we
calculate these hopping rates using Eqs. \eqref{b1} - \eqref{b3}. These values
are functions of both the temperature of the solid and the barrier against
diffusion, which we approximate as 50\% and 70\% of the desorption energy,
$E_\mathrm{D}$, of the relevant adsorbate-substrate pair for surface and bulk
diffusion, respectively.  For O$_2$-O$_2$, we have a measured value of
$E_\mathrm{D}^\mathrm{O_2}=7.85\times10^{-2}$ eV from an experimental study by
\citet{acharyya2007}.  Lacking other data, one could make the crude assumption
that atomic oxygen desorption is $\sim$ 0.5 times this value, and ozone
desorption as 1.5 times this value, giving
$E_\mathrm{D}^\mathrm{O}=3.92\times10^{-2}$ eV and
$E_\mathrm{D}^\mathrm{O_3}=1.18\times10^{-1}$ eV for O and O$_3$, respectively.
Unfortunately, we were unable to find much other previous work on O-O$_2$ or
O$_3$-O$_2$ desorption energies; however, \citet{cuppen2007} estimate these
values to be $E_\mathrm{D}^\mathrm{O}=4.74\times10^{-3}$ eV and
$E_\mathrm{D}^\mathrm{O_3}=1.03\times10^{-2}$ eV for O and O$_3$ adsorbed on an
O$_2$ substrate.  Here, we assume that the values derived from the work of
\citet{acharyya2007} represent a rough upper limit, while those in
\citet{cuppen2007} represent a lower limit, and use the average of the two.
This gives values of $E_\mathrm{D}^\mathrm{O}=2.15\times10^{-2}$ eV for O and
$E_\mathrm{D}^\mathrm{O_3}=6.41\times10^{-2}$ eV for O$_3$. B99 assumed that
the concentration of atomic oxygen in their ice remained very low throughout
the experiment, given its reactivity with molecular oxygen. We find that, in
our models at 5 K, values for the O-O$_2$ diffusion barriers above
$E_\mathrm{B}^\mathrm{O}\sim2.60\times10^{-2}$ eV lead to a large buildup of
atomic oxygen in the ice which is most likely unphysical. Values lower than
this, such as the one used here, result effectively in a constant near-zero
concentration of atomic oxygen in the ice, as it reacts quickly with
surrounding species. 

\begin{figure}[htb]
  \includegraphics[height=5.5cm]{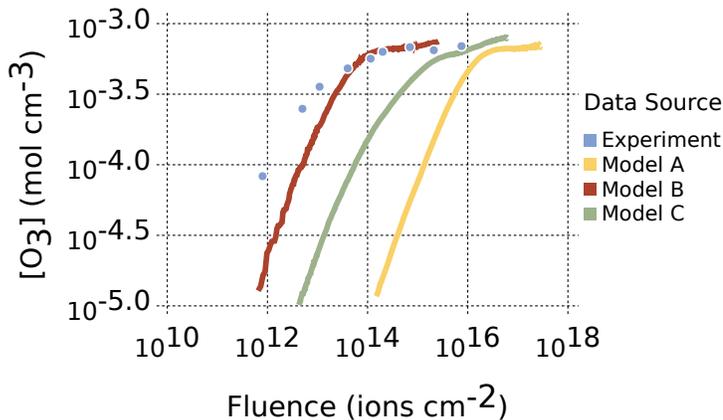}
  \caption{Ozone abundances both as calculated by \texttt{CIRIS} and from B99. }
  \label{f5}
\end{figure}

Plotted in Fig. \ref{f5} are ozone abundances as a function of proton fluence
for the experimental data of B99, as well as several calculated abundance
curves. Shown in yellow are data from Model A, in which the distances between
proton collisions are calculated on-the-fly based on the magnitude of the
energy-dependent cross-sections, as given by Eq.  \eqref{track-mfp}.  A
comparison between the experimental data and this model shows that, in both,
ozone abundances increase at roughly the same rate before reaching
$\sim7\times10^{-4}$ mol cm$^{-3}$.  After abundances have reached this quasi
steady-state value, further irradiation results in only marginal increases in
ozone.  The fluence at which such an inflection in growth occurs, however, is
much larger in the model than measured in B99. This discrepancy is likely due
to the differing ice thicknesses considered.  Since the ices used in the
experimental study were roughly two orders of magnitude thicker than the
$\sim0.1 \;\mu$m ice considered in this work, each proton which hit the surface
deposited less energy in our base model simulation. To illustrate this effect,
we include simulations in Fig. \ref{f5},Models B and C, in which the distance
between proton collisions was artificially reduced from our calculated value of
$\sim300$~\AA, to 3.3 and 33 \AA, respectively.  These data show that, as more
energy per proton is transferred to the solid, the theoretical results approach
the experimental values. From Fig.  \ref{f5}, one can also see that the best
agreement between our model and the experiment is obtained when $\Delta x =
3.3$~\AA. Given the thickness of our simulated ice, this means that the proton
collides $\sim300$ times before exiting the system.  If one used the Model A
results for protons of $\sim300$~\AA$\;$ between collisions, our model would
predict an ice thickness of $\sim10^5$~\AA, or 10 $\mu$m, approximately equal
to the actual thickness of the ice studied in B99. 

The chemistry behind the ozone abundances given in Fig. \ref{f5} begins with
the arrival of the first proton. The secondary electrons which form along the
track of this primary ion are the main drivers of atomic oxygen production
through both dissociative attachment with O$_2$ and dissociative recombination
with O$_2^+$. Atomic oxygen is mainly depleted through reaction with O$_2$,
resulting in the formation of the ozone which begins to build up in the ice.
This reaction between triplet O and triplet O$_2$ continues to be the primary
formation route for ozone throughout the simulation. Once formed, ozone is
destroyed through two main non-thermal pathways which occur as a result of
continued irradiation. The first of these is through direct ionization by both
the primary ion and secondary electrons, resulting in O$_3^+$, which is
destroyed via the charge neutralizations with either secondary electrons or
anions, which occur at the end of each track calculation phase of the code. The
other main destruction mechanism for ozone is reaction with singlet atomic and
molecular oxygen, both of which are formed either through direct excitation or
as a product of ion-ion and ion-electron recombinations.

\section{Summary and Conclusions}

In this work, we have presented a new stochastic model which calculates the
physical and chemical changes of solids exposed to high-energy protons.
Beginning with pure O$_2$ ice we are able to follow, on a
collision-by-collision basis, the tracks of energetic protons and secondary
electrons. The reactive species, such as atomic oxygen, which form as a result
of the non-thermal processes induced by the irradiation cause significant
chemical changes in the ice, even at the very low temperatures considered in
this work where thermal diffusive chemistry is inefficient. As shown in Fig.
\ref{f5}, we are able to reproduce both the quasi-steady-state abundances of
ozone in such an irradiated material, as well as predict the approximate
thickness of the ice studied by B99.

Though our ultimate goal is to better understand the degree to which
interactions between cosmic rays and interstellar grain ice mantles contribute
to the formation of complex molecules, the techniques presented here are not
specific to the chemistry of these extreme environments, and can be applied to
other systems as well. 
One area of practical interest where \texttt{CIRIS} can compliment experimental work is 
in modeling irradiated polymers, such as poly(vinyl chloride) or PVC.
These materials are widely used in environments in which they are exposed to 
ionizing radiation \citep{nicholson2006}. Previous experimental work has investigated
PVC radiolysis \citep{laverne2008}, and our code could compliment such studies by providing a means to 
investigate possible radiolysis pathways, or to simulate the system under 
physical conditions and exposures not practical or obtainable in the laboratory.
Ionizing radiation also poses serious health and safety risks for any
future missions to Mars and the Moon, and our code could be used to
evaluate potential radiation shielding materials, such as those 
composed of polymers. This type of shielding is generally more effective at stopping 
energetic ions than metals, but poses a greater risk of structural 
failure due to prolonged exposure  or flammability, in part, from 
radiolysis products like gaseous H$_2$ \citep{nasa2006,laverne2008}.
By allowing for simulations of the physiochemical evolution of potential 
shielding materials, use of \texttt{CIRIS} could aid mission planners in selecting the safest,
most effective solutions.

A major consideration for future applications to other
systems is the availability, or lack thereof, of relevant experimental data. In
calculating tracks and collisional probabilities for each neutral species
considered in this work, our code draws on cross-sectional data for electron
and proton impact ionization and excitation. For the chemistry, we rely on
desorption energies to calculate hopping rates, and kinetic studies to
determine the branching fractions. Unfortunately, there is often significant
uncertainty in desorption energies, as is the case for O-O$_2$ binding. For the
kinetics of charge recombination processes such as ion-ion and ion-electron
reactions, the unique properties of storage rings allow for experiments, such
as the one reported by \citet{zhaunerchyk2008}, that can furnish exactly the
kind of data which can be directly used in our code, i.e.  dissociation
pathways, the electronic states of products, and branching fractions.
Though we have had to rely mostly on previous gas-phase studies in modeling the 
comparatively simple O$_2$ ice system, this approach is not equally applicable for all systems. 
For instance, \citet{ribeiro2015} irradiated CH$_3$CN ice and observed 
products with notably different abundances from what would be expected in the gas phase, likely because of 
the influence of the condensed medium on the non-thermal ion and electron induced
chemistry. Such experimental work is of significant value when considering irradiation chemistry on 
a microscopic level and future applications should focus on systems where solid state 

Nevertheless, this model may represent a first-step towards a better
understanding of complex, interconnected phenomena. Future versions of our code
will focus on improvements that will enable us to examine other aspects of the
irradiation of a solid in more detail, despite the promising results reported
here of our simulations of the O$_2$ ice system. First, we would like to
consider displacement of lattice species due to elastic collisions in more
detail.  Improving these aspects will allow us to better model the changes to
the ice lattice caused by collisions between bulk species and the primary ion,
as well as to begin examining additional processes associated with ion-solid
collisions, such as sputtering, in which bound species on or near the surface
of the target material are released into the gas phase. Another aspect of the
model that is important to continue developing is electron transport. This
aspect is treated in more detail by the \texttt{CASINO} \citep{hovington1997}
and \texttt{PENELOPE} \citep{salvat2006} models, and it may be possible to
incorporate the theory they employ into future versions of \texttt{CIRIS}.
Related to this, electrostatic effects such as Coulombic forces could be
incorporated that could allow for a better picture of the motion of charged
species in the solid. 
Improvements such as these will enable us to better investigate
irradiation induced non-thermal desorption mechanisms. In addition to sputtering, these improvements include desorption induced by
electronic transitions (DIET), Auger stimulated ion desorption (ASID), 
and electron stimulated ion desorption (ESID) \citep{ribeiro2015}.
These processes are of particular astrochemical interest, because they represent
a means by which the complex molecules formed on interstellar dust grain surfaces can be

C.N.S. wishes to thank the Advanced Research Computing Services (ARCS) center
at the University of Virginia for use of the RIVANNA supercomputer. E.H.
wishes to thank the National Science Foundation for continuing to support the
astrochemistry program at the University of Virginia. This research has made
use of NASA's Astrophysics Data System Bibliographic Services.


\bibliography{bibliography} 
\bibliographystyle{rsc} 

\end{document}